\documentclass[%
reprint,
superscriptaddress,
showpacs,
nofootinbib, nobibnotes, 
 amsmath,
 amssymb,
 aps,
 prl
]{revtex4-1}

\usepackage{microtype}
\usepackage{blindtext}
\usepackage{dcolumn}
\usepackage{bm}
\usepackage{color, soul, colortbl} 
\usepackage[colorlinks=true,
						linkcolor=blue,
						urlcolor=blue,
						citecolor=blue,
						bookmarks=true,
						pdfborder={0 0 0}]{hyperref}

\usepackage{xcolor}
\usepackage{mdframed} 

\usepackage{graphicx,epsfig}
\usepackage{subfigure}
\usepackage{amsmath,amsfonts}
\usepackage{xfrac}
\usepackage{enumerate}
\usepackage[overload]{empheq}
\usepackage{mathtools}
\usepackage{cases} 
\usepackage{cancel} 

\begin{document}
\title{Group interactions modulate critical mass dynamics in social convention}

\author{Iacopo Iacopini}
\affiliation{Department of Network and Data Science, Central European University, 1100 Vienna, Austria}
\affiliation{Aix Marseille Univ, Universit\'e de Toulon, CNRS, CPT, Marseille, 13009, France}
\author{Giovanni Petri}
\affiliation{Mathematics and Complex Systems Research Area, ISI Foundation, Via Chisola 5, 10126 Turin, Italy}
\author{Andrea Baronchelli}
\thanks{These authors jointly supervised this work}
\affiliation{Department of Mathematics, City, University of London, EC1V 0HB, London, United Kingdom}
\affiliation{The Alan Turing Institute, British Library, 96 Euston Road, NW12DB, London, United Kingdom}
\author{Alain Barrat}
\thanks{These authors jointly supervised this work}
\affiliation{Aix Marseille Univ, Universit\'e de Toulon, CNRS, CPT, Marseille, 13009, France}
\affiliation{Tokyo Tech World Research Hub Initiative (WRHI), Tokyo Institute of Technology, Tokyo, Japan}


\begin{abstract}
How can minorities of individuals overturn social conventions? The theory of critical mass states that when a committed minority reaches a critical size, a cascade of behavioural changes can occur, overturning apparently stable social norms. Evidence comes from theoretical and empirical studies in which minorities of very different sizes, including extremely small ones, manage to bring a system to its tipping point. Here, we explore this diversity of scenarios by introducing group interactions as a crucial element of realism into a model for social convention. We find that the critical mass necessary to trigger behaviour change can be very small if individuals have a limited propensity to change their views. Moreover, the ability of the committed minority to overturn existing norms depends in a complex way on the group size. Our findings reconcile the different sizes of critical mass found in previous investigations and unveil the critical role of groups in such processes. This further highlights the importance of the emerging field of higher-order networks, beyond pairwise interactions.
\end{abstract}

\maketitle

\section*{Introduction}

The theory of critical mass argues that apparently stable social conventions can be overturned by a minority of committed individuals if such minority reaches a critical size \cite{granovetter1978threshold, schelling2006micromotives, xie2011social}. In this view, the power of small factions comes not from their authority or wealth but from the commitment to the cause and also, crucially, from their size. Evidence from different contexts however shows a wide range of possible sizes. On the one hand, several studies have found that rather large minority sizes were required to overturn a majority. Qualitative analyses of gender conventions in corporate leadership roles have hypothesised that a critical mass of $30\%$ of the population is necessary in order for the tipping point to be reached \cite{kanter1977some, dahlerup1988small}. Related observational work has proposed a higher critical mass size approaching $40\%$ of the population \cite{grey2006numbers}. Controlled experiments of social coordination have brought empirical evidence for tipping points in the dynamics of social conventions, finding a critical threshold of $25\%$ of the population~\cite{centola2018experimental}. On the other hand, numerous observations suggest that even a minority counting just tens of committed individuals, and not significant fractions of the population, may trigger abrupt social and normative change. Social movements \cite{diani1992concept} offer several examples in this sense (see \cite{gladwell2010small, chenoweth2011civil, farmer2019sensitive} and examples therein), and a data-driven analysis of linguistic norm change in English and Spanish pointed that committed minorities as small as $0.3\%$ of the population can impose their view~\cite{amato2018dynamics}.\newline

To understand the origin and nature of critical masses, several frameworks have been proposed to investigate their role in human behaviour, the dynamics of opinions, and the emergence of norms and consensus, starting from simple models inspired by statistical physics\cite{castellano2009statistical,vespignani2012modelling,sen2014sociophysics, baronchelli2018emergence,davis2020phase}. In particular, the naming game (NG) model has been influential in the theoretical description of the emergence of social norms \cite{baronchelli2006sharp,dall2006nonequilibrium,baronchelli2007nonequilibrium}. The model describes how a shared convention can emerge in a population of agents that interact locally with their peers, without any central coordination. It has brought theoretical support to the tipping points hypothesis, as a critical mass of $10\%$ was shown to be able to induce norm change in this model~\cite{xie2011social}. Subsequent generalisations of this model yielded critical mass varying between $10\%$ and $40\%$ of the population depending on the strength of individual commitment \cite{niu2017impact}. Adding further ingredients to the model, it was even possible to obtain vanishing sizes for the critical masses \cite{zhang2012analytic,thompson2014propensity,mistry2015committed,pickering2016analysis,doyle2017effects}. However, despite the wide range of observed critical mass sizes and of potential theoretical descriptions, little attention has been devoted to the problem of explaining how the initial group building up the critical mass itself emerges and what are its structural and dynamical determinants. Qualitative hypothesis include the existence of an intermediate phase composed by an initially small core of committed individuals able to recruit the missing mass of peers necessary to trigger a behavioural cascade \cite{centola2013homophily}. To this aim, small groups could exploit the role of high-resource individuals to mobilise the masses \cite{marwell1993critical}, homophily and local coalition formation \cite{centola2013homophily}, or the role of non-committed individuals sitting at the periphery of the social network (‘slacktivists’) \cite{barbera2015critical}. In all cases, the effect of small groups--and group interactions--can be determinant. Indeed, group interactions are the building blocks of real-world social systems, from discussions in real life between a group of friends to collaboration networks, and including online discussions on online social media and forums, which can involve large numbers of individuals~\cite{atkin1974mathematical, freeman1980q, patania2017shape, benson2018simplicial, cencetti2021temporal}. Recent works in complex systems research have focused on taking into account these more realistic higher-order (non-pairwise) interactions \cite{lambiotte2019networks, battiston2020networks, porter2020nonlinearity, bick2021higher, benson2021higher,torres2021and,battiston2021physics}. This includes models of opinion dynamics such as the majority rule \cite{galam2013modeling,noonan2021dynamics}, or extensions of other models describing social and evolutionary dynamics to hypergraphs~\cite{alvarez2020evolutionary, neuhauser2020multibody, hickok2022bounded, guo2021evolutionary,neuhauser2021consensus}, leading to important changes of dynamical behaviour. For instance, group interactions can dramatically alter social contagion dynamics and lead to a rich phenomenology including abrupt transitions, bi-stability and critical mass phenomena~\cite{iacopini2019simplicial}. 

Here, we investigate the role of group interactions on critical mass effects. To this aim, we extend the widely adopted naming game framework, which has been shown to explain the outcome of controlled experiments \cite{centola2018experimental}, to account for more realistic interaction patterns and dynamics. We improve the modeling along three directions. First, we propose a model to describe norm evolution upon group interactions between any number of agents. We thus encode these many-body interactions into the hyperlinks of a hypergraph~\cite{berge1984hypergraphs}, which provides a more faithful representation of real-world social structures~\cite{battiston2020networks, torres2021and}, and define new rules for group agreement. Second, we take into account that social influence is in general not perfect \cite{kelley1957certainty, fuegen2004intensity, au2011strength}. Thus, individuals may successfully interact with one another without necessarily converging on a norm adoption as a result of the interaction. More specifically, while in the standard naming game model a successful coordination is followed by a certain and exclusive adoption of the norm that allowed the coordination~\cite{baronchelli2006sharp, baronchelli2018emergence} (perfect social influence \cite{axelrod1997dissemination}), here individuals may be reluctant to let go of alternative conventions even when they successfully manage to coordinate with one another on a specific norm. Third, we inform the model with a variety of real-world data concerning the structure of empirical social networks and their microscopic (non-pairwise) interactions. This represents an  improvement with respect to standard all-to-all, pairwise, or synthetic approaches. Taken together, these three advances contribute towards a more realistic representation of both social interactions and dynamics. Extensive numerical simulations of this model on empirical data and synthetic hypergraphs show that the critical mass required to induce norm change is dramatically reduced when non-committed members of the population are not fully susceptible to social influence. We also find a rich phenomenology in which groups modulate the takeover by helping sustaining the view of the minority. In particular, we unveil a non-monotonic dependency in the long term dynamical output: interactions in very small or very large size turn out to be more favourable to the committed minority than interactions in groups of intermediate size. Our results hold when considering data from very different social contexts and in a simplified version of the interaction structure, for which we develop an analytical mean-field approach, allowing us to get further insights and study the interplay between social influence and size of the committed minority.

\section*{Results}
\subsection*{The Framework}\label{sec:model}

In order to include a more realistic description of social interactions, we generalise the standard NG by considering that agents can interact not only in pairs but also in groups of arbitrary size. Groups are indeed the most natural units through which individuals engage with each other in social contexts~\cite{atkin1974mathematical, freeman1980q, patania2017shape, cencetti2021temporal}, and it has been shown that considering these higher-order interactions may reveal a rich phenomenology \cite{battiston2020networks,battiston2021physics}. While the usual NG considers that the agents are located on the nodes of a network and interact along its links, we therefore encode here the group interactions between the agents as the hyperlinks of a hypergraph. This higher-order representation of the social structure, in which pairwise interactions (links) $[i,j]$ are called 1-interactions, expands the more traditional representation offered by graphs by considering relationships between any number of agents, such as 2-interactions (triangles) $[i,j,k]$, 3-interactions (tetrahedra), etc: a hyperlink is simply a set of nodes $[p_0,p_1,\dots,p_{k-1}]$ that conveniently represents a multi-body interaction between $k$ nodes~\cite{berge1984hypergraphs, battiston2020networks}. 

The evolution rules of the game in the generalised setting are the following. At each time step a hyperlink $e$ is randomly chosen and a { speaker} agent is chosen at random among the nodes composing $e$. All the other nodes participating in the same hyperlink act as { hearers}. The speaker selects a random name, say $A$, from its current vocabulary and communicates it to the group. In the pairwise NG interactions, there is only one hearer and the agreement can be reached if and only if the hearer has the name $A$ in its vocabulary. Here, the introduction of group interactions requires the definition of a new generalised condition on the group of hearers for an agreement to be possible. Multiple options are obviously possible. The strictest possible condition is that the agreement can be reached by the group only if all the hearers have $A$ in their vocabularies. This agreement rule can be seen as a { unanimity condition} rule between the hearers' vocabularies. On the other hand, the weakest condition would be that an agreement can be reached provided at least one of the hearers knows the name already. We call this the { union} rule, as it is enough that the name is in the union of the vocabularies of the hearers. The unanimity condition seems more suitable when modeling consensus rule, as it implies that all members of the interacting group need to know a name to converge to it. On the other hand, in the union rule, an alliance of two agents (the speaker and one hearer) can make a group of arbitrary size converge. We thus focus mainly on the unanimity condition rule in the main text. We show the results for the union rule in the Supplementary Note 2, and mention where relevant how the results differ between the two rules. Intermediate rules such that a given fraction or number of hearers need to have the name in their vocabularies could also be considered. 

\begin{figure}
	\centering
	\includegraphics[width=.96\linewidth]{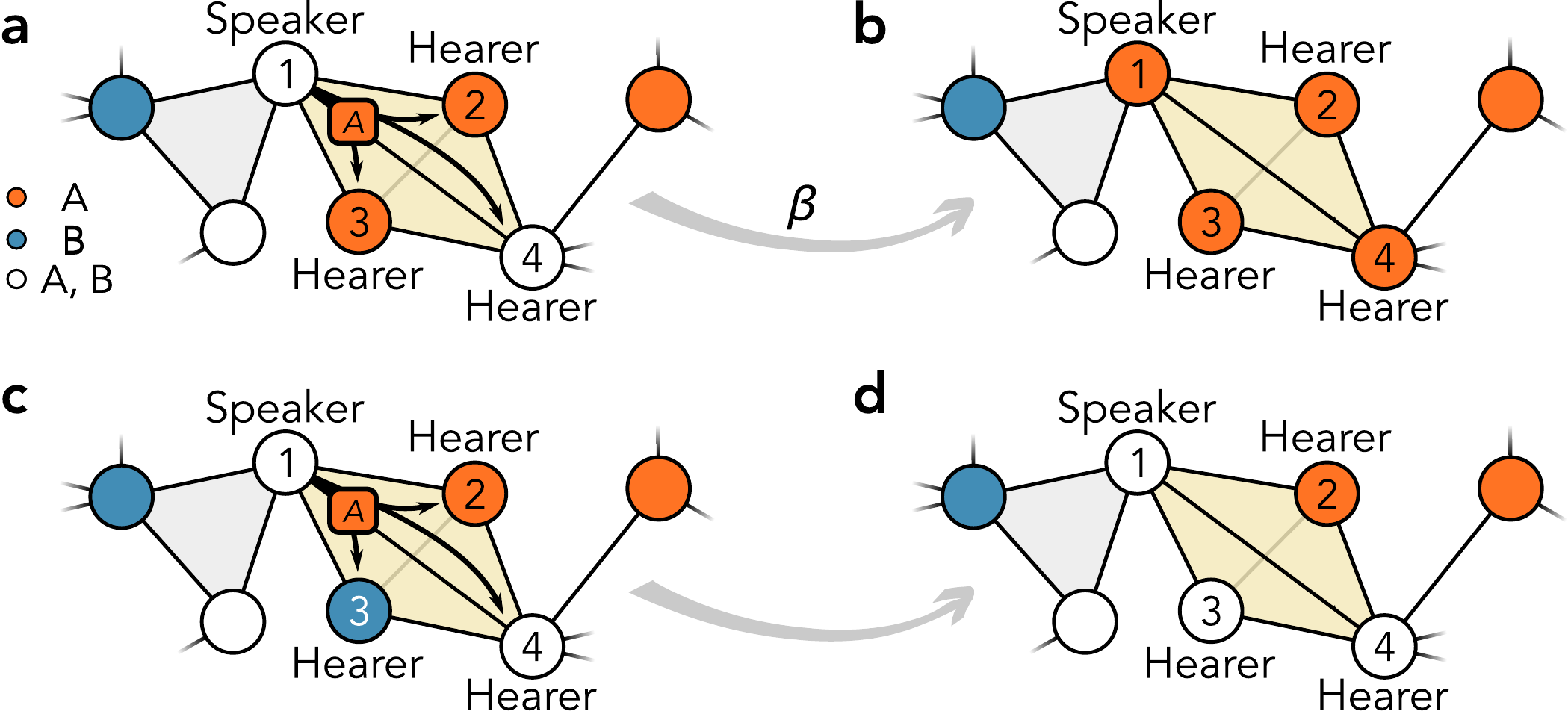}
	\caption{{\bf Dynamics of the model.} Agents are represented by the nodes of a social structure composed by interacting groups of different sizes. The vocabulary of the agents--for simplicity containing at most only two names (or conventions) $\{A, B\}$--is reflected in the colours of the nodes as shown in the legend. At each interaction a group is chosen at random (highlighted in yellow in the figure) together with a speaker (node 1), while the remaining nodes act as hearers.
	Here we illustrate the { unanimity} rule (see model definition).
	({\bf a}) The speaker chooses a name at random from its vocabulary (here, $A$), and communicates it to the rest of the group. Since $A$ is present in the vocabularies of all the hearers (nodes 2 and 3 support $A$, while node 4 knows both names), the group can reach an agreement.
	({\bf b}) With probability $\beta$ the group agrees on the chosen name, and all nodes involved immediately update their vocabulary to $A$, erasing $B$. With probability $1-\beta$ instead the agreement does not happen.
	({\bf c}) In this case, the speaker selects $A$, but node 3 does not possess $A$ in its vocabulary. 
	({\bf d}) Thus, there cannot be agreement in the group. Nevertheless, all hearers update their vocabularies by adding the heard name, i.e., node 3 switches from $A$ to $A, B$. 
	}\label{fig:model}
\end{figure}

In all cases, the propensity of the hearers to accept the convergence to a consensus in the group is controlled by a parameter $\beta\in[0,1]$ (notice that this parameter contributes only when an agreement is possible)~\cite{baronchelli2007nonequilibrium}. Thus, if an agreement can be reached, two possibilities exist [Fig.~\ref{fig:model}a]: ({i}) with probability $\beta$ all the nodes of the considered hyperlink agree on the chosen name $A$ and erase all the other names from their vocabularies [Fig.~\ref{fig:model}b]; ({ii}) with probability $1-\beta$ there is no convergence but the nodes who did not have $A$ add it to their vocabulary. When agreement is instead not possible, all nodes who did not have $A$ add it to their vocabulary [Fig.~\ref{fig:model}c,d]. Thus, the parameter $\beta$ modulates social influence, i.e. the propensity of individuals to change their behaviour to meet the demands of a social environment. The smaller the $\beta$ the less the individuals participating are prone to change their views in spite of the social influence mechanism~\cite{jenness1932role, myers1971enhancement, judd2010behavioral}. 

Finally, we allow for the presence of a committed minority among the agents. The dynamics of these committed agents does not obey to the aforementioned rules. Instead, this fraction $p$ of agents always sticks to the same name and do not change it nor updates their vocabulary \cite{xie2011social}. 
For simplicity of notations and consistency throughout the manuscript, we assign to these agents the name $A$ (and we  denote their fraction as $A_c$). We will also denote with $n_x(t)$ the fraction of agents supporting name $x$ at a given time $t$, and with $n_x^*$ the corresponding values in the stationary states reached in the long time regime. 

\subsection*{Critical mass dynamics}

\begin{figure*}
	\centering
	\includegraphics[width=1\linewidth]{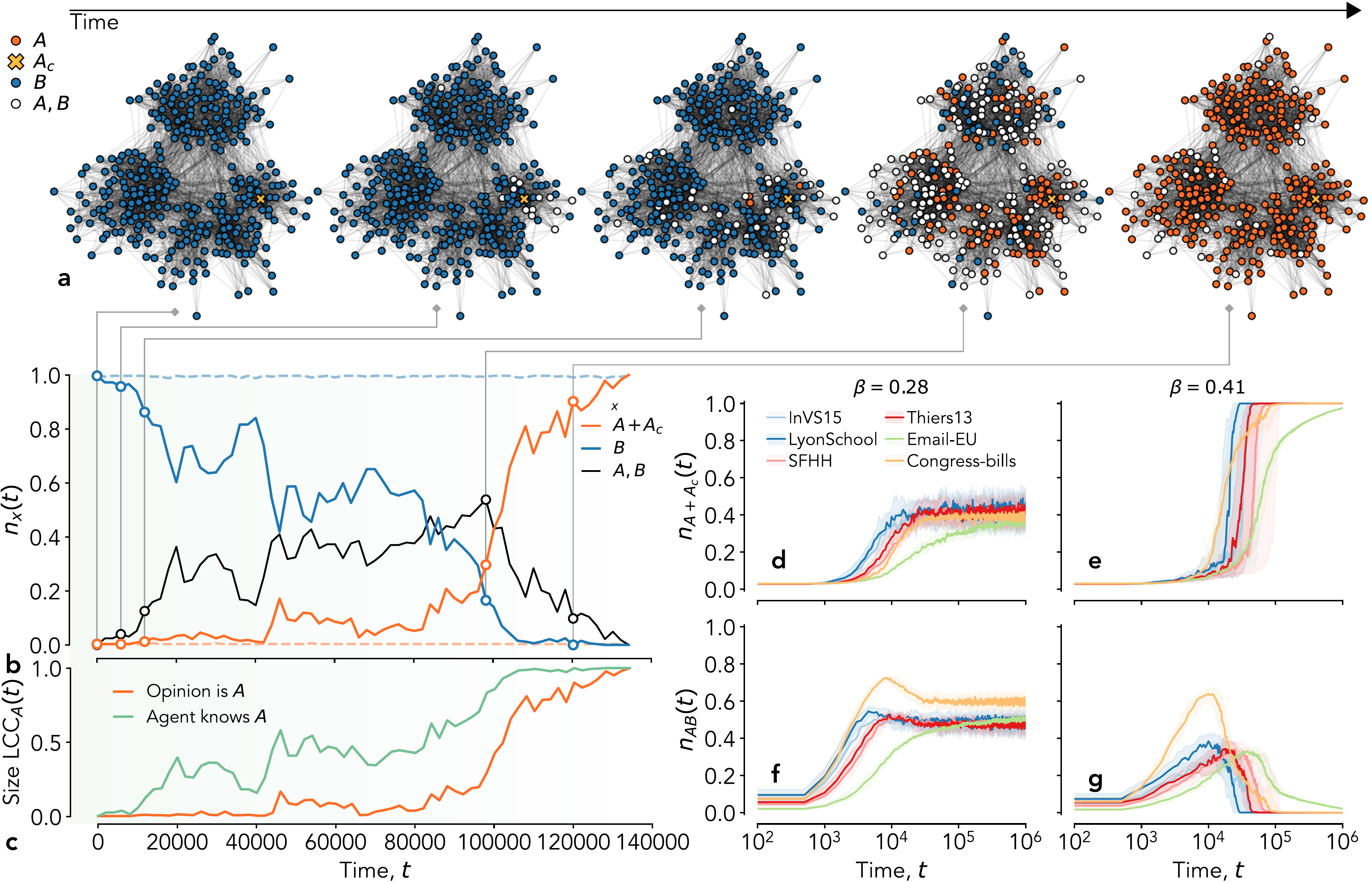}
	\caption{{\bf Critical mass dynamics.}
	({\bf a}) Illustrative example of a simulation of the 
	Naming Game (NG) with unanimity rule on an empirical social structure (Thiers13), where a minority $A_c$ of one single committed individual supporting $A$--consisting of 0.3\% of the population of 327 individuals--overturns the stable social norms and reaches global consensus (under imperfect communication, with social influence parameter $\beta=0.336$).
	({\bf b}) Temporal evolution of the fraction $n_x(t)$ of nodes supporting name $x$. Different solid lines correspond to different names, $x=\{A+A_c, B, (A, B)\}$. Dashed lines are reported as a benchmark, representing the case with perfect communication ($\beta=1$).
	({\bf c}) Temporal evolution of the normalised size of the largest connected component (LCC) of nodes supporting name $A$ (red curve) and nodes that have $A$ (but not necessarily $A$ only) in their vocabulary (green curve).
	({\bf d,e,f,g}) Temporal evolution of the dynamics with committed minorities ($p=3\%$) on empirical higher-order structures.
	The social structures are constructed from empirical data sets collected in six different context: a workplace ({InVS15})~\cite{genois2015data}, a primary school ({LyonSchool})~\cite{stehle2011high}, a conference ({SFHH})~\cite{isella2011s}, a high school ({Thiers13})~\cite{mastrandrea2015contact}, email communications ({Email-EU})~\cite{paranjape2017motifs} and a political congress ({Congress-bills})~\cite{fowler2006legislative}.
	The temporal evolution of the densities of nodes holding name $A$ and holding both $A$ and $B$ are reported in panels ({\bf d,e}) and ({\bf f,g}), respectively, for two different values of the parameter $\beta$ quantifying the efficacy of reaching an agreement within a group, namely $\beta=0.28$ ({\bf d,f}) and $\beta=0.41$ ({\bf e,g}). The results over different runs of stochastic simulations are reported as median values (solid lines) and values contained within the 25th and 75th percentiles (shaded areas).
}\label{fig:committed_takeover}
\end{figure*}

We first show how the model allows for minority takeover even in extreme cases. In Fig.~\ref{fig:committed_takeover}a-c we show an example of how a small minority, consisting here of a single committed individual (0.3\% of the population of 327 individuals), can overturn the majority. The example reports the results of a single simulation (representative of $95\%$ of the runs we have simulated) of a NG with the unanimity rule and $\beta=0.336$. The considered empirical social structure consists of face-to-face interactions--as recorded by wearable sensors in a French high-school~\cite{mastrandrea2015contact}--and includes group interactions of sizes ranging from 2 to 5 (see { Methods} for details). The committed individual is selected at random. We show in Fig.~\ref{fig:committed_takeover}a visualisations of the structure of interactions (showing for simplicity only the links and not the groups of larger sizes) and of the states of the nodes at several times during the simulation. Figure \ref{fig:committed_takeover}b moreover shows the temporal evolution of the fraction of nodes supporting a given name $x$ (solid lines) until the absorbing state with all nodes converging on $A$ is reached. The circular markers on the curves correspond to the times at which the five configurations of Fig.~\ref{fig:committed_takeover}a were observed. In these visualisations, we distinguish the committed agent by a lighter orange cross ($A_c$). Notice that this is the only node initially supporting $A$ ($n_A(0)=0$), while the remaining nodes are initially assigned the name $B$ ($n_{B}(0)=1-p$), represented by a blue colour. As time evolves, the committed agent starts to spread name $A$ locally: The fraction of nodes supporting both names at the same time (but without preference for either one, and represented by white nodes) increases around the committed one, who converts a number of blue neighbours into white nodes, that subsequently start to further diffuse the name $A$. At later times, we observe a change in the slope of $n_{A}(t)$: the fraction of nodes having adopted $A$ starts to rapidly increase up to the point when $n_{A,B}(t)$ reaches a maximum and starts decreasing, while the initial minority takes over.

Figure \ref{fig:committed_takeover}b shows also, as a benchmark, the outcome of a simulation with the same committed individual but $\beta=1$. The associated temporal evolution, where the minority remains as such (dashed lines), confirms the central role played, in the minority takeover, by the parameter $\beta$ that encodes the propensity of individuals to accept the convergence to a consensus in a group. Finally, Fig.~\ref{fig:committed_takeover}c shows the temporal evolution of the normalised sizes of the largest connected component (LCC) of nodes supporting name $A$ (or $A_c$) and of the LCC of nodes that know $A$, namely with vocabulary either $A$, $A_c$ or $A, B$. More precisely, we consider the subgraph induced by the nodes having the considered status, we compute the size of its LCC and then normalise it by the total number of nodes. As the simulation evolves the subgraphs start growing around the committed node until they span the entire population, the subgraph of nodes knowing $A$ naturally growing faster than the subgraph of nodes whose inventory contains only $A$.

\subsection*{Analysis of Different Regimes}

\begin{figure*}
	\centering
	\includegraphics[width=0.9\linewidth]{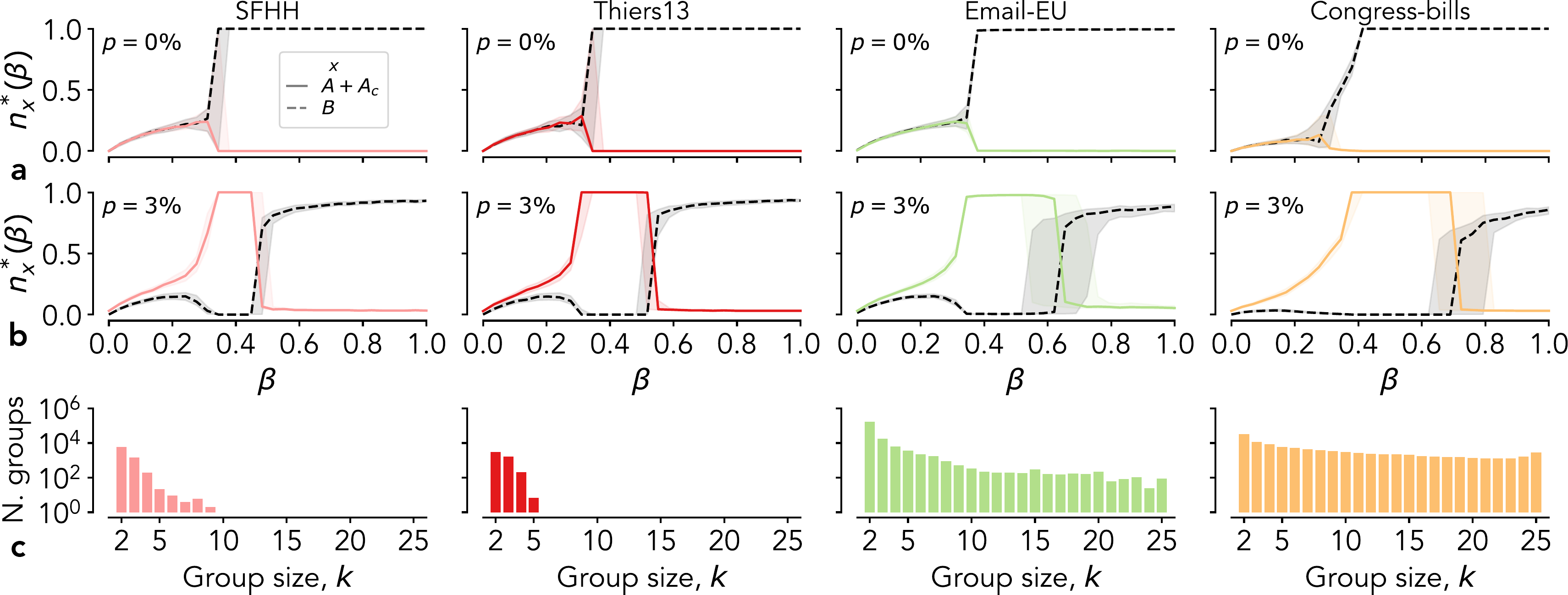}
	\caption{{\bf Stationary state of the naming game (NG) dynamics with { unanimity rule} simulated on empirical higher-order structures, and histograms of the group sizes of these structures.}
	The considered social structures correspond to empirical data sets collected in four different contexts (see {Methods}): a conference (SFHH), a high school (Thiers13), email communications (Email-EU) and a political congress (Congress-bills).
	Panels ({\bf a}) and ({\bf b}) correspond respectively to simulations without and with committed minorities $A_c$ supporting name $A$
	(respectively, fraction of committed $p=0$ and $p=3\%$). In these panels, we plot the fraction of nodes supporting name $x$ in the stationary state, $n^*_x(\beta)$, obtained by means of numerical simulations on each data set, as a function of
	the social influence parameter $\beta$. Lines (continuous and dashed, respectively associated to names $A+A_c$ and $B$) and shaded areas refer to the median values and values contained within the 25th and 75th percentile measured over 150 runs.
	The panels ({\bf c}) show the histograms of group sizes associated to each data set, where a group of size $k$ represents a higher-order interaction (of order $k-1$) between $k$ nodes.
}\label{fig:sim_empdata}
\end{figure*}

In order to clarify the role of the underlying social structure and of the corresponding group interactions, we study the evolution of the model on  different empirical data sets of interactions. To this aim, we rely on publicly available data sets of different kinds of temporally-resolved social interactions, from which we can construct aggregated higher-order (group) representations (see also \cite{iacopini2019simplicial}). The data we use were collected in six different contexts, namely a workplace (InVS15), a primary school (LyonSchool), a conference (SFHH), a high school (Thiers13, the data set used in Fig. \ref{fig:committed_takeover}), email communications (Email-EU) and a political congress (Congress-bills). The resulting hypergraphs are very different in terms of group size distribution, with some containing groups up to size 24. We provide additional information on the data and the data aggregation methodology in { Methods}.\newline 
We first simulate the model on each structure, with $p=3\%$ (see { Methods} for details of the numerical simulations) and different values of $\beta$. Figure~\ref{fig:committed_takeover}d-g shows the results averaged across 50 different runs for each empirical structure. In each panel, the fraction of nodes supporting name $x$ is plotted as a function of time [panels (d,e) $x=A+A_c$, panels (f,g) $x=A,B$] for two different values of $\beta$, namely $\beta=0.28$ (d,f) and $\beta=0.41$ (e,g) [see also Supplementary Figure 1]. Clearly, and despite the significant differences among the data sets, the dynamics falls into the same two radically different regimes depending on the parameter $\beta$. For low values of $\beta$ [$\beta=0.28$, Fig.~\ref{fig:committed_takeover}d,f], all the curves stabilise after some time--that depends on the considered structure--on similar and intermediate values of the densities of individuals with norm $A$ or $B$. In this scenario the initial very small minority of committed individuals manages to strongly expand the reach of the name $A$, but does not convince the entire population. Instead, there is a co-existence regime, where the number of agents supporting name $A$ remains globally fairly constant, while, microscopically, the nodes continue to switch between different states. We note that this regime is not present in the traditional NG ($\beta=1$) where a global consensus is always reached at long times. 
The picture changes for larger values of $\beta$ [$\beta=0.41$, Fig.~\ref{fig:committed_takeover}e,g]. After an initial transient, there is an abrupt transition (with a temporal scale that depends on the particular structure) after which the minority manages to convince the entire population and simulations reach the absorbing state of global coherence (with name $A$). Notice how the $n_{AB}(t)$ curve typically presents a peak right before $n_{A+A_c}(t)$ starts to increase, corresponding to the behaviour already highlighted in the single run investigated in Fig.~\ref{fig:committed_takeover}b: the information about name $A$ diffuses from the committed, making the vocabulary of individuals who have $B$ first become $(A,B)$ and finally switch to $A$. We also illustrate this behaviour in Supplementary Figure 2, by showing that the average time for individuals to adopt $A$ (exclusively) for the first time increases with their distance from a committed individual.

We further characterise the behaviour of the model by studying how the stationary state changes with $\beta$. Simulation results are shown in Figure~\ref{fig:sim_empdata}, where we plot the fraction of nodes holding each name ($A$ or $B$) in the stationary state as a function of $\beta$, either without committed individuals ($p=0$) or with a fraction $p=0.03$ of committed (see Supplementary Figures 3-4 for the results for the other data sets, as well as for different values of the fraction of committed $p$ and for the union condition rule for group agreement). Figure~\ref{fig:sim_empdata}a-b highlights the rich behaviour of the model as $\beta$  changes from $0$ to $1$, with several clearly distinct regimes. In the absence of committed minorities [$p=0$, Fig.~\ref{fig:sim_empdata}a], we initialise the simulations with 40\% of individuals supporting name $A$ ($n_B(0)=0.6$), and we obtain two distinct regimes. For low values of $\beta$, the system reaches a stationary state  where the two names co-exist and have on average the same density: neither one dominates, and a substantial fraction of individuals hold the two names in their vocabulary (as even when there is no local convergence because $\beta$ is small, the hearers add the speaker's norm to their vocabulary, which favours the emergence of agents having both names). However, as consensus becomes easier (increasing $\beta$), the social influence within groups tends to favour the name of the initial majority ($B$). Above a critical value $\beta_c$ that varies with the structure, the dynamics falls into the absorbing state with no $A$ agents left. In contrast with the pairwise NG model in \cite{baronchelli2007nonequilibrium}, the transition is not abrupt. There is in fact a third intermediate regime--whose extent strongly depends on the social structure--where names co-exist, but $B$ dominates. 

A radically different scenario emerges when a committed minority is present. The system  exhibits three different regimes, illustrated in Figure~\ref{fig:sim_empdata}b that gives the results of simulations performed with a small seed of committed agents (chosen at random) with name $A$, all other agents having initially the name $B$ ($p=0.03$, $n_A(0)=0$, $n_B(0)=1-p$). At small values of $\beta$, as for the former case, local convergence within each group interaction is hard, and a stationary state with co-existence of the two names is reached. Name $A$, despite being known initially only to a small minority, reaches nevertheless a substantial fraction of the population and is actually more represented than $B$ even in this regime. As $\beta$ increases, the advantage gained by $A$ becomes rapidly stronger until we reach an absorbing state in which the initial minority wins and conquers the entire population, while name $B$, which was initially shared by the majority ($97\%$) of the population,  disappears. This regime persists for a certain range of $\beta$ values that we call $\Delta\beta^*$. At larger $\beta$ finally, we enter a third regime where the committed minority is not able to spread its norm widely, and the system converges to a stationary state where the name supported by the initial majority prevails, with the obvious exception of the committed agents, and with a small fraction of agents in contact with the committed who tend to have a shared vocabulary $(A,B)$.

This phenomenology is qualitatively robust, despite different underlying social interaction structures.  Fig.~\ref{fig:sim_empdata}c shows the numbers of groups of each size contained in each data set. Face-to-face interactions involve relatively low number of agents, so that group sizes are limited; on the contrary, email communications (Email-EU) and political networks of bills co-sponsoring (Congress-bills) can correspond to larger groups, involving up to 24 individuals, with a heterogeneous distribution of group sizes. The social structure influences the quantitative results of the long term dynamics, as the larger intervals $\Delta\beta^*$ are associated to the data sets that include large groups. We have verified that these quantitative differences are due indeed to group size distributions and not to particular correlations that might be present in the data, by performing simulations using a group-size-based mean-field approach, as in Supplementary Figure 5: in these simulations, the data are reshuffled so that the group size distribution is preserved but other correlations are destroyed. The differences in the width of the $\Delta\beta^*$ interval in which the committed minority prevails are preserved, showing that the size of the groups plays an important role. We investigate this role more systematically in the next section.

\subsection*{The Role of Group Size}

\begin{figure}
	\centering
	\includegraphics[width=1\linewidth]{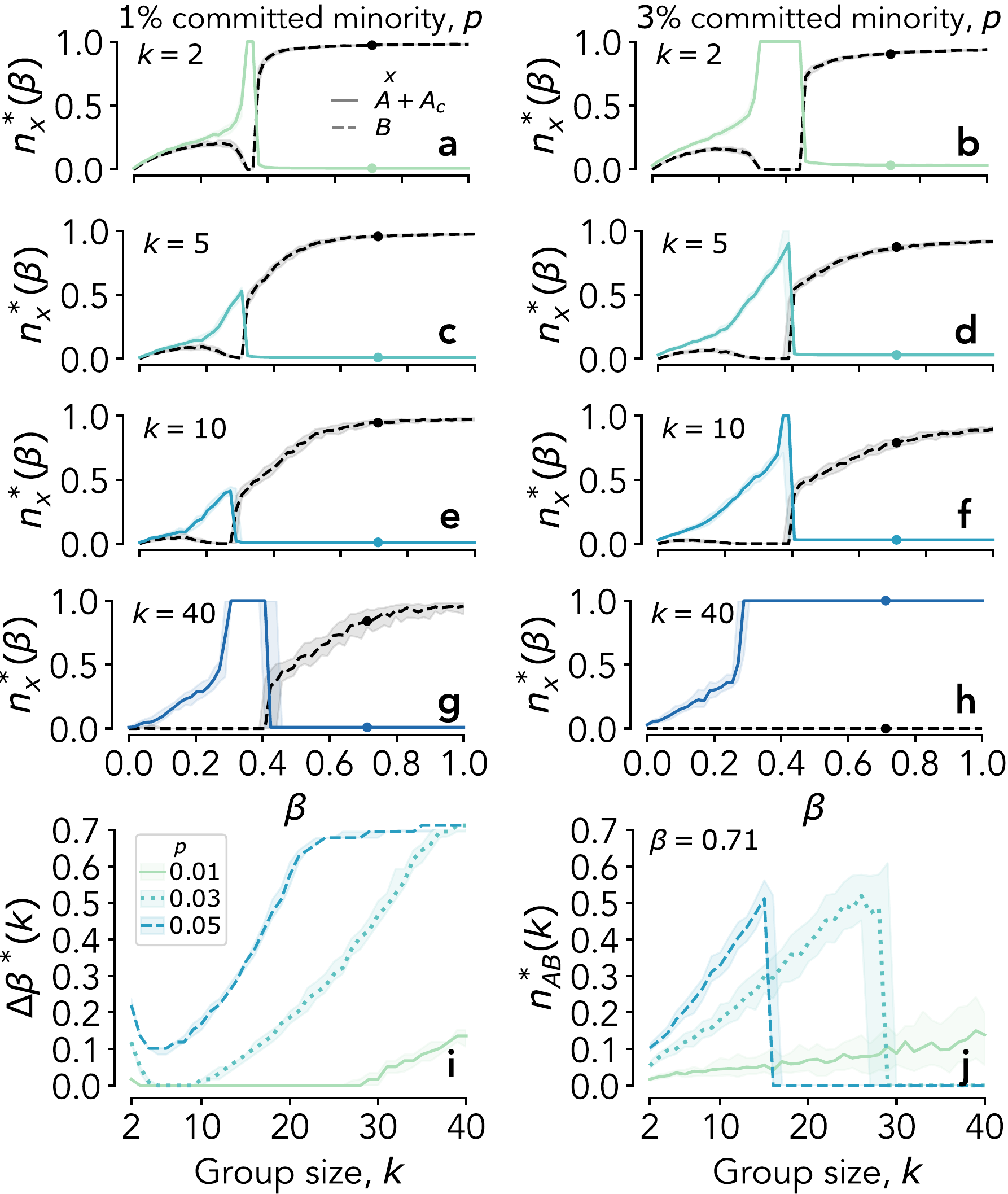}
	\caption{{\bf Higher-order (group) effects in the naming game (NG) for different values of the social influence parameter $\beta$.} We consider 
	$(k-1)-$uniform hypergraphs, i.e. regular structures in which each interaction involves exactly $k$ agents. Group agreement follows the unanimity rule. 
	({\bf a,b,c,d,e,f,g,h}) The density of nodes supporting name $x$ in the stationary state, $n^*_x(\beta)$, obtained by means of numerical simulations, is shown as a function of
	the social influence parameter $\beta$ ($A_c$: committed minority). Lines (continuous and dashed, respectively associated to names $A+A_c$ and $B$) and shaded areas refer to the median values and values contained within the 25th and 75th percentiles of the 50 numerical simulations. Panels ({\bf a,c,e,g}) and ({\bf b,d,f,h}) correspond to simulations with different sizes of committed minorities supporting name $A$, namely fractions $p=0.01$ and $p=0.03$ of the population. In the initial state, all the other agents hold norm $B$. Four different group sizes are considered: $k=2$ ({\bf a,b}), $k=5$ ({\bf c,d}), $k=10$ ({\bf e,f}) and $k=40$ ({\bf g,h}).
	The range $\Delta\beta^*$ of $\beta$ values for which $n^*_A=1$ (i.e., the committed minority manages to convert the whole population), is plotted in ({\bf i}) as a function of the group size $k$ and for different values of the fraction $p$ of committed (see legend). 
	We show in Supplementary Figure 8 the equivalent of this panel for the union rule.
	({\bf j}) Fraction of nodes $n^*_{AB}$ holding both names in the stationary state as a function of $k$ for different values of $p$.
}\label{fig:sim_groupsize}
\end{figure}

The results discussed above show a similar phenomenology for different social contexts corresponding to different interaction patterns, with different group sizes and generalised degree distributions~\cite{courtney2016generalized} (see Supplementary Figure 6). In this section, we explore more systematically the impact of group size on the outcome of the dynamics, and in particular its interplay with the existence of a committed minority. We therefore consider synthetic interaction structures where we control the distribution of group sizes. For simplicity, we consider homogeneous structures in which individuals interact in groups of fixed size (uniform hypergraphs). Figure~\ref{fig:sim_groupsize} shows the results of stochastic simulations on such $(k-1)$-uniform hypergraphs, where each hyperlink consists of exactly $k$ nodes. In Fig.~\ref{fig:sim_groupsize}a-h we plot the median fraction of nodes supporting name $x$ in the stationary state, $n^*_x(\beta)$, as a function of $\beta$ for several group sizes. Panels (a,c,e,g) and (b,d,f,h) show the results of simulations initiated with a different fraction of committed nodes holding name $A$, set respectively to $p=0.01$ and $p=0.03$ (all the other nodes holding initially name $B$). When varying the group size from $k=2$ (a,b) to $k=5$ (c,d), the range $\Delta\beta^*$ in which the committed convert the whole population to $A$ decreases. However, when we further increase the group size (g,h), $\Delta\beta^*$ increases again. Thus, increasing $k$ changes the range of $\beta$ values for which the minority manages to take over the entire population in a non-monotonic way. This range also depends on the committed population size $p$, with a broader range for larger $p$. Similar plots for different agreement rules can be found in Supplementary Figure 7. Fig.~\ref{fig:sim_groupsize}i further highlights this non-monotonic dependence of $k$ by plotting the optimal range $\Delta\beta^*$ as a function of $k$ for different values of $p$, for the unanimity rule. We also show in Supplementary Figure 8 the evolution with $k$ of the minimal and maximal values of this range, $\beta^*_{min}$ and $\beta^*_{max}$, for both rules. For the intersection rule, both are non-monotonic with $k$: at small $k$, $\beta^*_{min}$ increases and $\beta^*_{max}$ decreases, while both reverse their tendency at larger $k$. On the other hand, for the union rule, $\beta^*_{max}$ increases monotonically with $k$ and only $\beta^*_{min}$ keeps a non monotonic behaviour. This complex behaviour can be interpreted by several competing effects of the group size.  Let us first consider large $\beta$. For the intersection rule, the unanimity condition makes it more difficult for an agreement to be possible on a minority name when the group size increases; on the other hand, more individuals are converted to the new name upon each single successful agreement. The competition between these effects leads to a non-monotonic $\beta^*_{max}$. For the union rule, the first effect does not exist and increasing group size makes it more probable for a minority speaker to find a hearer knowing the  minority name: $\beta^*_{max}$ increases monotonically with $k$. At small $\beta$ and low $k$, co-existence of norms is favoured for both rules by increasing $k$ (hence $\beta^*_{min}$ increases), because it is more probable that both norms are represented when a larger group is chosen to interact; as $\beta$ is small, local consensus is not very probable and instead the most probable outcome is that all agents gain the word of the speaker in their inventories, thus becoming $(A,B)$. As $k$ continues to increase however, the local convergence of large groups starts to dominate and $\beta^*_{min}$ starts thus to decrease. We note that similar competing effects of group sizes have been described in a different type of cascade dynamics \cite{petri2018simplicial}, leading to non-monotonic group size effects. Moreover, this phenomenology provides quantitative support to the so-called group size paradox well-studied in sociology~\cite{oliver1988paradox}. We also notice that even for large values of $\beta$ (in the regime where the name supported by the committed minority does not reach the entire population) group interactions still play a major role in keeping the system out of the absorbing state of the majority and promoting the alternative norm. This is evident from Fig.~\ref{fig:sim_groupsize}j, where we report the fraction of agents having both names in the stationary state as a function of $k$, for a fixed value of $\beta=0.71$, which--depending on $k$ and $p$--might fall inside or outside of the optimal range $\Delta\beta^*$. The corresponding points are also highlighted in Fig.~\ref{fig:sim_groupsize}a-h. We see that $n_{AB}^*$ increases with the group size, up to the point where the minority wins. Notice how this fraction of individuals holding both norms becomes easily much larger than $p$, highlighting how the committed individuals propagate the knowledge of the alternative name to a substantial part of the population, even when they do not manage to completely reverse the initial majority.

\subsection*{Analytical approach}

To get further insights on the observed behaviour, its dependency on the parameters of the model, as well as to access the impact of the microscopic rule for group agreement, we finally consider a mean-field (MF) description of the generalised model. We limit here our study to groups of $3$ nodes as the number of potential combinations becomes soon intractable as the group size increases. We thus consider an infinitely large homogeneously mixing population such that each interaction involves three nodes taken at random, obeying the rules of the NG defined above, under the { unanimity} condition rule for reaching an agreement. The equations for the dynamical evolution of the densities of agents having each possible vocabulary, $n_{A}$, $n_{B}$ and $n_{AB}$, can then be written as:
\begin{subequations}\label{eq:MF_int}
	\begin{align}
		d_t n_A &= -2n_A^2n_B+\left(\frac{5}{2}\beta-1\right) n_A^2 n_{AB}  \nonumber \\
		& -2n_An_B^2-3n_An_Bn_{AB} \nonumber \\
		&+(4\beta-1)n_An_{AB}^2+\frac{3}{2}\beta n_{AB}^3 +\frac{5}{2}\beta p^2 n_{AB}  \nonumber \\
		&+p[ - 2 n_An_B+(5\beta-1) n_A n_{AB}+4\beta  n_{AB}^2] \\
		d_t n_B &= -2n_B^2n_A+\left(\frac{5}{2}\beta-1\right) n_B^2 n_{AB}  \nonumber \\
		&-2n_Bn_A^2-3n_Bn_An_{AB} \nonumber \\
		&+(4\beta-1)n_B n_{AB}^2+\frac{3}{2}\beta n_{AB}^3 -2p^2n_B \nonumber \\
		&- p[4n_An_B+2n_B^2+3n_Bn_{AB}] \\
		n_{AB} &= 1-n_A-n_B-p
	\end{align}
\end{subequations}
where $d_t$ denotes the time derivative. 

Similar equations can be derived for different agreement rules, such as for the less strict { union} rule, in which having one of the hearers knowing the name expressed by the speaker is potentially enough to reach consensus in the entire group. The evolution equations in this case read
\begin{subequations}\label{eq:MF_union}
	\begin{align}
		d_t n_A &= 2(\beta-1)n_A^2n_B+(\frac{5}{2}\beta-1)n_A^2n_{AB} \nonumber \\
		&-2 n_An_B^2 +(6\beta-3)n_An_Bn_{AB} \nonumber \\
		&+(4\beta-1)n_An_{AB}^2 +3\beta n_Bn_{AB}^2+\frac{3}{2}\beta n_{AB}^3\nonumber \\
		&+ p^2[2\beta n_B +\frac{5}{2}\beta n_{AB}] + p[2(2\beta-1)n_An_B  \nonumber \\
		&+ (5\beta-1)n_An_{AB}+6\beta n_B n_{AB}+4\beta n_{AB}^2] \\
		d_t n_B &= 2(\beta-1)n_B^2n_A+(\frac{5}{2}\beta-1)n_B^2n_{AB} \nonumber \\
		&-2 n_Bn_A^2 +(6\beta-3)n_Bn_An_{AB} \nonumber \\
		&+(4\beta-1)n_Bn_{AB}^2 +3\beta n_An_{AB}^2+\frac{3}{2}\beta n_{AB}^3 \nonumber \\
		&-2p^2n_B -p[2n_B^2+(3-3\beta)n_Bn_{AB}]\nonumber \\
		&-p[-2\beta n_{AB}^2+4n_An_B] \\
		n_{AB} &= 1-n_A-n_B-p   \  .
	\end{align}
\end{subequations}
Equations for the benchmark pairwise rule, without higher-order interactions, are reported in Supplementary Note 1.

To explore the interplay between the size of the committed minority and the social influence parameter $\beta$ in determining the final state of the system evolving according to these equations, we integrate them numerically. For both rules, we use the same initial condition as in the previous sections: we fix a fraction $p>0$ of committed agents and a value for $\beta$, and we set $n_A(0)=n_{AB}(0)=0$ so that initially the only agents with name $A$ are the committed ones, all the others holding name $B$ ($n_B(0)=1-p$). We record the evolution of the densities $n_x(t)$ until they reach a stationary value (results are also confirmed by directly imposing the stationarity condition $d_t = 0$, and solving the resulting equations through the program { Mathematica}).

This allows us to study how the final fraction of agents holding name $A$, $n^*_{A+A_c} \equiv n_A(t\to\infty)+ p$, depends on $p$ and $\beta$. Results are shown in Fig.~\ref{fig:MF} as three-dimensional phase diagrams. We also show, to make the link with previous figures clearer, a curve (in black) of $n^*_{A+A_c}$ as a function of $\beta$ for a fixed fraction of committed individuals ($p=0.08$). With the { unanimity} condition [Fig.~\ref{fig:MF}a] three regimes emerge, as observed before with stochastic simulations: at low $\beta$ the committed manage to create a co-existence of names, with $n^*_{A+A_c} >> p$, but do not overturn the initial majority. In the central region, whose width depends on the size of the committed minority $p$ (see also Supplementary Figure 9), the new name $A$ prevails and the initially general name $B$ completely disappears. Finally, at large enough $\beta$ the system is not much perturbed by the committed. Overall, the larger the committed minority, the higher $\beta$ needs to be to avoid the initial name to disappear. This phenomenology persists for different agreement rules, as for the less restrictive { union} condition shown in Fig.~\ref{fig:MF}b. In this case, for $p=0.08$ the third regime vanishes and the optimal range $\Delta\beta^*$ expands up to the case $\beta=1$ corresponding to the perfect social influence scenario (as for the standard NG). We also note that the results of our MF approach are in perfect agreement with the ones from stochastic simulations, here reported as white circles (further comparisons with stochastic simulations are reported in Supplementary Figures 10-11). Additional heatmaps for both rules and the pairwise benchmark are reported in Supplementary Figures 12-13.

\section*{Discussion}

Online connectedness is reportedly speeding up the process of collective behavioural change \cite{kooti2012emergence, centola2015spontaneous} through the adoption of new norms \cite{centola2018experimental, becker1999constructing, bicchieri1999great, del2016spreading}. In this scenario, clarifying the microscopic mechanisms driving this process is key to gain a better understanding of our society and to design possible interventions aimed at contrasting undesired effects. At the same time, understanding how policy can create tipping points where none exist and how it can push the system past the tipping point are fundamental questions whose answer might change the way in which we address major societal challenges \cite{nyborg2016social}, such as accelerating the post-carbon transition \cite{farmer2019sensitive} or contrasting vaccine-hesitancy \cite{chevallier2021covid,iten2013p037}.

The model for tipping points dynamics in social convention introduced in this work extends the usual naming game framework towards more realism. First, we moved beyond peer-to-peer communication by considering group interactions not only restricted to pairs, as many actual interactions, both in real life and online, involve group discussions. In addition, we investigated the effects of imperfect social influence and considered real social structures for interactions. Our results show that critical mass dynamics can be initiated by minorities of very different sizes, including by an (almost) arbitrarily small minority, and that groups play a crucial role in determining the minority takeover. For example, one single individual with no special power or wealth can overturn the social conventions held by a group of hundreds of peers. Counter-intuitively, this happens when agents are not fully inclined to let go of the convention they currently use in favour of a new one. Note that if agents are extremely reluctant to change, the system remains in a state where both norms co-exist at a similar level and many agents hold both norms. In order to consider realistic interactions, we have performed extensive numerical simulations on higher-order social networks constructed from empirical data and including group interactions of very different sizes, confirming that the findings hold in a broad region of the parameter space and for different interaction settings.
Moreover, we have unveiled an interesting role of the size of interacting groups. Recently, large groups have been found to have a dominant role in seeding and sustaining contagion processes on hypergraphs \cite{st2021influential}. Here, we have shown a non-monotonic behaviour in which both small and large group interactions favour the committed minority. We believe that our group communication model, which includes higher-order effects in micro-interactions, opens up a new direction in naming game applications to study opinion spreading and norm emergence. For instance, it would be relevant to explore how the influence of a minority can be maximized \cite{kempe2003maximizing, kitsak2010identification, st2021influential}, depending on whether its members tend to be cohesive or part of different groups. More broadly, this confirms the relevance of going beyond network representations and taking into account higher-order interactions when modelling social phenomena \cite{battiston2020networks,battiston2021physics}.

It is important to delimit the scope of our findings. The major limitation of our results is, of course, that they have been obtained in the context of a single theoretical model, i.e., the naming game framework. However, this model has been previously used in several theoretical and empirical studies on tipping point dynamics in social convention, successfully reproducing the results obtained in controlled experiments~\cite{centola2018experimental}. As such, by generalising this model towards the inclusion of realistic interactions, by confirming how committed minorities of varying sizes can take over and even small committed minorities can be dramatically effective, and by shedding light on the effect of higher-order interactions, our work contributes to the development of more realistic modelling approaches and to the understanding of the critical mass phenomena. Naturally, it would be interesting to expand the results to other modelling frameworks of emergence or social cooperation \cite{galam2007role, galam2013modeling, cardillo2020critical}. An even more substantial step further would involve the design of novel controlled experiments \cite{centola2018experimental,deamicis2020understanding,poncela2016humans} to empirically assess the impact of the (different) group interactions introduced and of the various model parameters. For instance, $p$ can be tuned by giving specific instructions to some participants to act as committed, while the role of $\beta$ could be mimicked by artificial agents similar to bots \cite{bertrand2020impact,monsted2017evidence} that could prevent convergence with a tunable probability.

A second limitation is that our model cannot fully account for the complexity of real world interactions. While this is certainly true, it is worth stressing that we have considered more realistic interactions patterns than ever before in this context, or for that matter in many multi-agent models. Future work may further enrich this aspect by considering the effect of community structures \cite{FORTUNATO201075} and of temporal (higher-order) networks \cite{mistry2015committed, chowdhary2021simplicial, neuhauser2021consensus}. Moreover, the model and the generalisations we have considered lend themselves easily to many natural extensions: in particular, the parameter $\beta$ describing the modulation of social influence could depend on agents' properties, such as their centrality, or on the size of each interacting group. The activity of a group (probability to be selected at each time step) could also depend on its size. Co-evolution of the interaction structure and of the norms could also be introduced in the model~\cite{horstmeyer2020adaptive}. Finally, it would be very interesting to investigate the behaviour of our model when agents committed to different opinions are present~\cite{xie2012evolution}, with two distinct minorities pushing each a distinct new norm against an initial well established one.

\section*{Methods}
\subsection*{Data Description and Aggregation}

We build empirical hypergraphs by aggregating six different data sets of temporally resolved interactions. Four of these data sets, provided by the SocioPatterns collaboration~\cite{sociopatterns}, describe face-to-face interactions collected in different social contexts: a workplace ({InVS15})~\cite{genois2015data}, a primary school ({LyonSchool})~\cite{stehle2011high}, a conference ({SFHH})~\cite{isella2011s}, and a high school ({Thiers13})~\cite{mastrandrea2015contact}. Data from these experiments are initially aggregated by using a temporal window of 15 minutes and the maximal cliques within each temporal snapshot are retained (a similar procedure was used in Ref.~\cite{iacopini2019simplicial}). We then simply filtered the cliques by removing those that appeared only once, and finally used them to build each of the four empirical hypergraphs considered.
The other two data sets involve considerably bigger interactions, and it is evident from the group size distributions reported in Fig.~\ref{fig:sim_empdata}. The { Email-EU} data set refers to email communications from a European research institution~\cite{paranjape2017motifs},  where each node represent a different email address and each hyperlink involves the sender and the (multiple) recipients of each message (1-second resolution). Finally, the { Congress-bills} data set refers to legislative bills in the U.S. congress~\cite{fowler2006legislative}, where each node represent a person in the congress and hyperlinks join sponsors and co-sponsors of bills put forward in the House of Representatives and the Senate. No additional data processing has been performed on these last two data sets, as they already come in the form of simplices from Ref.~\cite{benson2018simplicial}.

The six empirical hypergraphs, in addition to describing very different types of interactions and contexts, are composed by different numbers of nodes $N$ and groups $E$: $N=217$, $E=3,704$ ({InVS15}), $N=242$, $E=8,010$ ({LyonSchool}), $N=403$, $E=7,741$ ({SFHH}), $N=327$, $E=4,862$ ({Thiers13}), $N=9,79$, $E=209,005$ ({Email-EU}), $N=1,718$, $E=105,929$ ({Congress-bills}). The histograms of group sizes are shown in Fig.~\ref{fig:sim_empdata} and in Supplementary Figure 3, while generalised degree distributions are reported in Supplementary Figure 6.

\subsection*{Stochastic simulations}
We run agent-based stochastic simulations of the generalised NG model on real-world social structures and on idealised homogeneous populations of $N=1000$ agents. In this latter case, simulations are performed assuming an homogeneous mixing population: all agents can potentially interact with each other. 
In both cases the dynamics evolves in the following way. At each timestep a group is chosen at random, either from the actual list of groups composing the empirical data set, or in the homogeneous case, by selecting at random $k$ nodes.
One of the nodes composing the selected group, randomly chosen, acts as a speaker and the remaining nodes as hearers. The status of each node is then updated according to the specific rules defined in the model. The process is repeated until the system reaches an absorbing state (with all the nodes holding the same norm) or a steady state for the densities of agents holding a given norm. Densities in the steady state are computed by taking the average over $100$ values sampled from the last 50,000 steps. The results shown in the figures correspond to median and standard deviations computed on 50 runs with random initial conditions (i.e., with a random selection of committed agents).

\subsection*{Data Availability}
The data sets are available from the original sources \href{http://www.sociopatterns.org/datasets/}{http://www.sociopatterns.org/datasets/} and \href{https://github.com/arbenson/ScHoLP-Data}{https://github.com/arbenson/ScHoLP-Data}. 

\subsection*{Code Availability}
The code is available at \href{https://github.com/iaciac/higher-order-NG}{https://github.com/iaciac/higher-order-NG}.

\subsection*{Author Contributions}
I.I., G.P., A. Baro., A. Barr. conceived and designed the study. I.I. wrote the code and performed the numerical simulations. I.I. and A. Barr. wrote the analytical equations. 
I.I., G.P., A. Baro., A. Barr. interpreted the results. 
I.I., G.P., A. Baro., A. Barr. wrote the manuscript.

\subsection*{Competing Interests}
The authors declare no competing interests.

\subsection*{Acknowledgements}
 I.I. and A.Barr. acknowledge support from the Agence Nationale de la Recherche (ANR) project DATAREDUX (ANR-19-CE46-0008). I.I. acknowledges support from the James S. McDonnell Foundation $21^{\text{st}}$ Century Science Initiative Understanding Dynamic and Multi-scale Systems - Postdoctoral Fellowship Award. A.Barr. also acknowledges support from Japan Society for the Promotion of Science (JSPS) KAKENHI Grant Number JP 20H04288. G.P. acknowledges support from Intesa Sanpaolo Innovation Center. The funder had no role in study design, data collection, and analysis, decision to publish, or preparation of the manuscript.


%


\clearpage
\setcounter{figure}{0}
\setcounter{table}{0}
\setcounter{equation}{0}
\makeatletter
\renewcommand{\theequation}{S\arabic{equation}}
\renewcommand{\thetable}{S\arabic{table}}
\renewcommand{\thesection}{Supplementary Note \arabic{section}}
\renewcommand{\figurename}{Supplementary Figure}

\setcounter{secnumdepth}{2} 

\widetext
\begin{center}
	\textbf{\large Supplemental Material:\\ Group interactions modulate critical mass dynamics in social convention}
\end{center}

\section{Mean field approach with different condition rules for group agreement}
In the main text we have restricted most of our attention to one particular rule for group agreement that we called {\it unanimity condition}. At the essence of this rule there is the requirement that the spoken name has to be present in all the vocabularies of the hearers. As such, we can mathematically refer to this condition as an ``intersection'' rule. Here, we complement the mean field description of the proposed model by describing, all together, a ``intersection'' agreement rule, a ``union'' agreement rule, and a pairwise case for comparison. 

As for the main text, we consider the simplest case in which there are only two names, $A$ and $B$, so that each vocabulary can have only three states, $A$, $B$, and $AB$. We further assume an homogeneous mixing population that interacts by means of 2-interactions, that is hyperedges of size 3 (triangles). In this case, given a speaker with vocabulary $S$ that communicates a name $A$ to two hearers with vocabularies $X$ and $Y$, the two agreement rules we have defined reduce to:

\paragraph*{Intersection rule:}
\begin{itemize}
	\item if  $A\in X\cap Y$:
	\begin{itemize}
		\item with probability $\beta$: 
		$S \Rightarrow \{A\}$, $X \Rightarrow \{A\}$, $Y \Rightarrow \{A\}$
		\item with  probability $1- \beta$: 
		$X \Rightarrow X \cup \{A\}$, $Y \Rightarrow Y \cup \{A\}$
	\end{itemize}
	\item if  $A\notin X\cap Y$:
	$X \Rightarrow X \cup \{A\}$, $Y \Rightarrow Y \cup \{A\}$
\end{itemize}

\paragraph*{Union rule:}
\begin{itemize}
	\item if  $A\in X\cup Y$:
	\begin{itemize}
		\item with probability $\beta$: 
		$S \Rightarrow \{A\}$, $X \Rightarrow \{A\}$, $Y \Rightarrow \{A\}$
		\item with  probability $1- \beta$: 
		$X \Rightarrow X \cup \{A\}$, $Y \Rightarrow Y \cup \{A\}$
	\end{itemize}
	\item if  $A\notin X\cup Y$:
	$X \Rightarrow X \cup \{A\}$, $Y \Rightarrow Y \cup \{A\}$
\end{itemize}

We denote with $n_{A}$, $n_{B}$ and $n_{AB}$ the fraction of agents in each state (omitting the temporal index for readability purposes), while $p$ denotes the fraction of agents committed to $A$. Considering both rules, the equations for the evolution of the fraction of agents in each state read:\newline

\paragraph*{``Intersection'' rule:}

\begin{subequations}\label{eq:SI:MF_int}
	\begin{align}
		d_t n_A &= -2n_A^2n_B+(\frac{5}{2}\beta-1) n_A^2 n_{AB}-2n_An_B^2-3n_An_Bn_{AB}+(4\beta-1)n_An_{AB}^2\nonumber \\
		&+\frac{3}{2}\beta n_{AB}^3 + \frac{5}{2}\beta p^2 n_{AB} +p[ - 2 n_An_B+(5\beta-1) n_A n_{AB}+4\beta  n_{AB}^2] \\
		d_t n_B &= -2n_B^2n_A+(\frac{5}{2}\beta-1) n_B^2 n_{AB}-2n_Bn_A^2-3n_Bn_An_{AB}+(4\beta-1)n_Bn_{AB}^2 \nonumber \\
		&+\frac{3}{2}\beta n_{AB}^3-2p^2n_B - p[4n_An_B+2n_B^2+3n_Bn_{AB}] \\
		n_{AB} &= 1-n_A-n_B-p
	\end{align}
\end{subequations}

\paragraph*{``Union'' rule:}

\begin{subequations}\label{eq:SI:MF_union}
	\begin{align}
		d_t n_A &= 2(\beta-1)n_A^2n_B+(\frac{5}{2}\beta-1)n_A^2n_{AB}-2 n_An_B^2 +(6\beta-3)n_An_Bn_{AB}\nonumber \\
		&+(4\beta-1)n_An_{AB}^2+3\beta n_Bn_{AB}^2+\frac{3}{2}\beta n_{AB}^3 + p^2[2\beta n_B +\frac{5}{2}\beta n_{AB}]\nonumber \\
		&+ p[2(2\beta-1)n_An_B + (5\beta-1)n_An_{AB}+6\beta n_B n_{AB}+4\beta n_{AB}^2] \\
		d_t n_B &= 2(\beta-1)n_B^2n_A+(\frac{5}{2}\beta-1)n_B^2n_{AB}-2 n_Bn_A^2 +(6\beta-3)n_Bn_An_{AB}\nonumber \\
		&+(4\beta-1)n_Bn_{AB}^2+3\beta n_An_{AB}^2+\frac{3}{2}\beta n_{AB}^3 \nonumber \\
		&-2p^2n_B-p[2n_B^2+(3-3\beta)n_Bn_{AB}-2\beta n_{AB}^2+4n_An_B] \\
		n_{AB} &= 1-n_A-n_B-p
	\end{align}
\end{subequations}

We also consider the pairwise case as a benchmark, whose equations are:\newline

\paragraph*{Pairwise case:}

\begin{subequations}\label{eq:SI:MF_pairwise}
	\begin{align}
		d_t n_A &= -n_An_B + \frac{1}{2}(3\beta-1)n_An_{AB}+\beta n_{AB}^2+\frac{3}{2}\beta p n_{AB}\\
		d_t n_B &= -n_An_B + \frac{1}{2}(3\beta-1)n_Bn_{AB}+\beta n_{AB}^2-pn_B\\
		n_{AB} &= 1-n_A-n_B-p
	\end{align}
\end{subequations}

\clearpage

\section{Supplementary results}

\begin{figure}[hb]
	\centering
	\includegraphics[width=\textwidth]{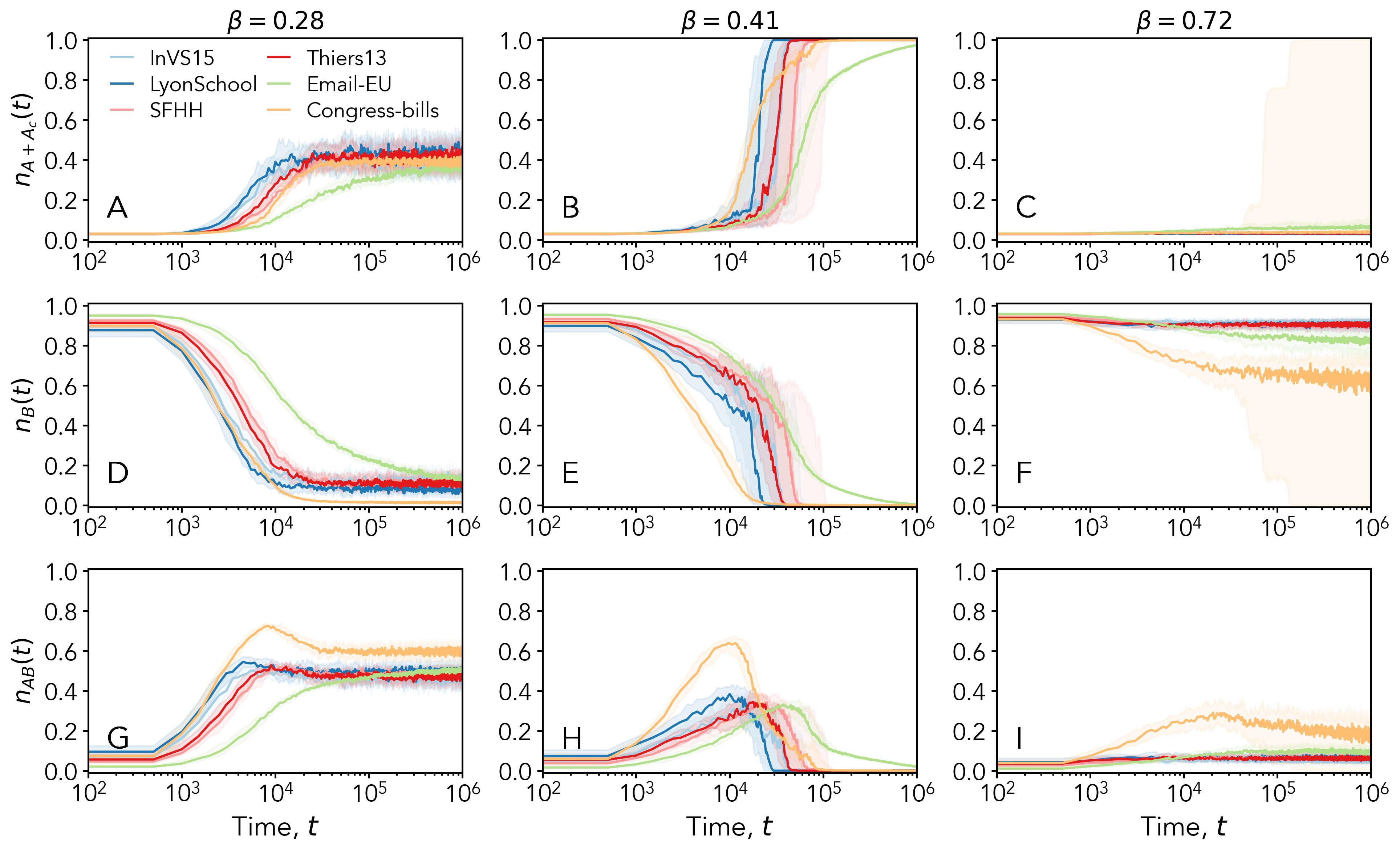}
	\caption{Temporal evolution of the dynamics with committed minorities on empirical higher-order structures.
		As for Fig.~2 of the main text, the considered social structures are constructed from empirical data sets collected in six different context: workplace ({InVS15}), a primary school ({LyonSchool}), a conference ({SFHH}), a high school ({Thiers13}), email communications ({Email-EU}) and a political congress ({Congress-bills}).
		The temporal evolution of the fraction of nodes supporting name $A$, $B$ and $A, B$ (with 3\% of committed individuals) is reported in panels ({\it A-C}), ({\it D-F}) and ({\it G-I}), respectively, for three different values of the communication efficiency, namely $\beta=0.28$ ({\it A,D,G}), $\beta=0.41$ ({\it B,E,F}) and $\beta=0.72$ ({\it C,F,I}). The results averaged over different runs of stochastic simulations are reported as solid curves and shaded areas, representing median values and values contained within the 25th and 75th percentiles.
	}\label{fig:SI:sim_tevo}
\end{figure}


\begin{figure}
	\centering
	\includegraphics[width=0.85\textwidth]{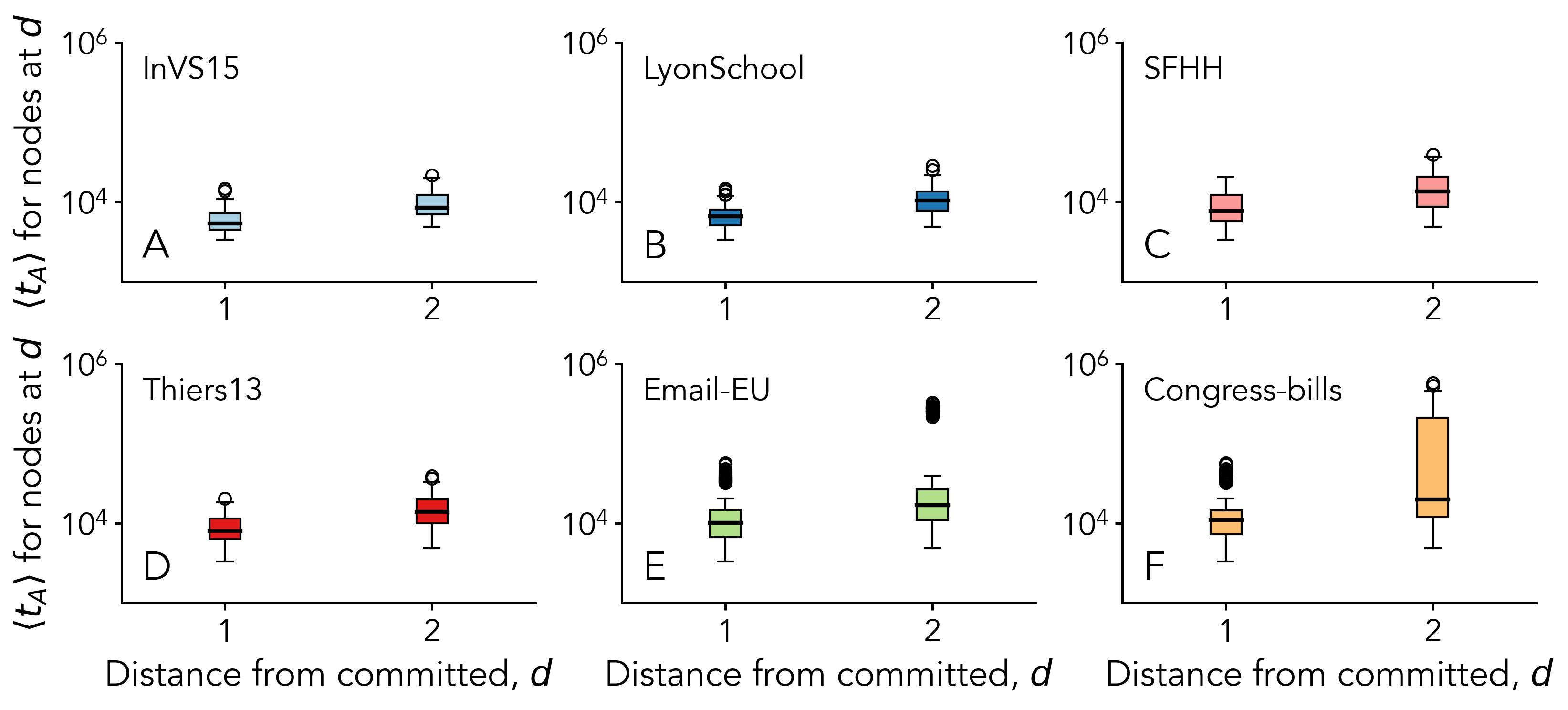}
	\caption{Adoption dynamics around committed individuals for simulations of the NG with the unanimity (``intersection'') condition for group agreement on empirical higher-order structures. The parameters are $p=0.03$ and $\beta=0.4$. The structures correspond to six different contexts: workplace ({InVS15}), a primary school ({LyonSchool}), a conference ({SFHH}), a high school ({Thiers13}), email communications ({Email-EU}) and a political congress ({Congress-bills}).
		For each data set, the boxplot shows the average time $\langle t_A \rangle $ that it takes for nodes at (graph) distance $d$ from committed individuals to switch to $A$. Notice that this differs from just having $A$ in the vocabulary, which instead has to be solely composed by $A$.}
\end{figure}


\begin{figure*}[t]
	\centering
	\includegraphics[width=\textwidth]{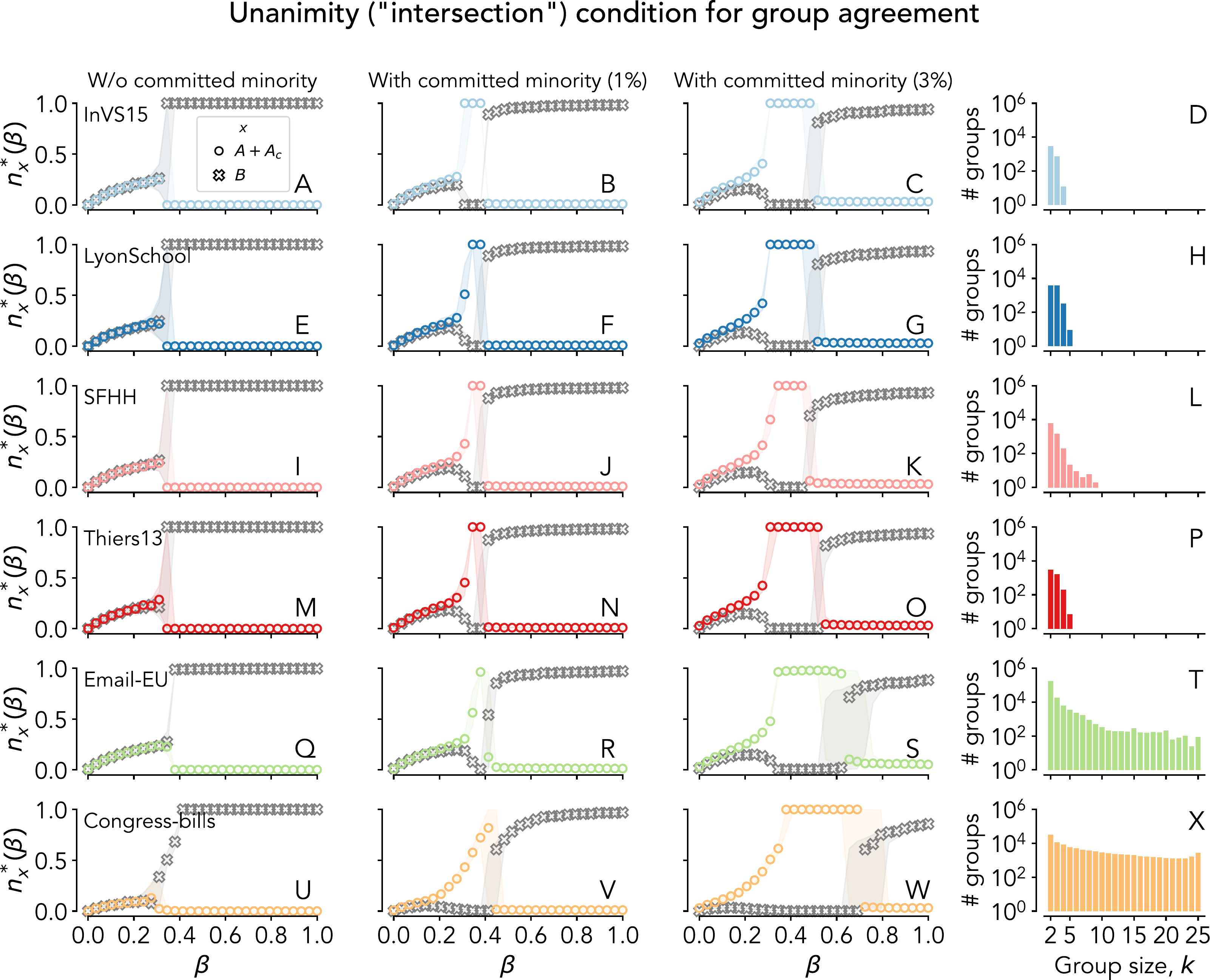}
	\caption{Simulations of the stationary state of the dynamics on empirical higher-order structures for a NG with the unanimity (``intersection'') condition for group agreement and associated group size frequencies.
		The considered social structures are the same of Fig.~2 of the main text and correspond to empirical data sets collected in six different context: workplace ({InVS15}), a primary school ({LyonSchool}), a conference ({SFHH}), a high school ({Thiers13}), email communications ({Email-EU}) and a political congress ({Congress-bills}). The phase diagrams in panels ({\it A,E,I,M,Q,U}), panels ({\it B,F,J,N,R,V}) and panels ({\it C, G, K, O, S, W}) correspond to simulation with a different fraction of committed minority, respectively set to $p=0$, $p=0.01$ and $p=0.03$. 
		The fraction of nodes supporting name $x$ in the stationary state $n^*_x(\beta)$, obtained by means of numerical simulations on each data set (row), is plotted as a function of the communication efficiency $\beta$.
		The results averaged over different runs of stochastic simulations are reported as points (circles and crosses, respectively associated to name $A$ and $B$) and shaded areas, representing median values and values contained within the 25th and 75th percentiles.
		Panels ({\it D,H,L,P,T,X}) show the group size frequencies associated to each data set, where a group of size $k$ represents a higher-order interaction between the $k$ nodes involved.
	}\label{fig:SI:sim_empdata_intersection}
\end{figure*}


\begin{figure*}[t]
	\centering
	\includegraphics[width=\textwidth]{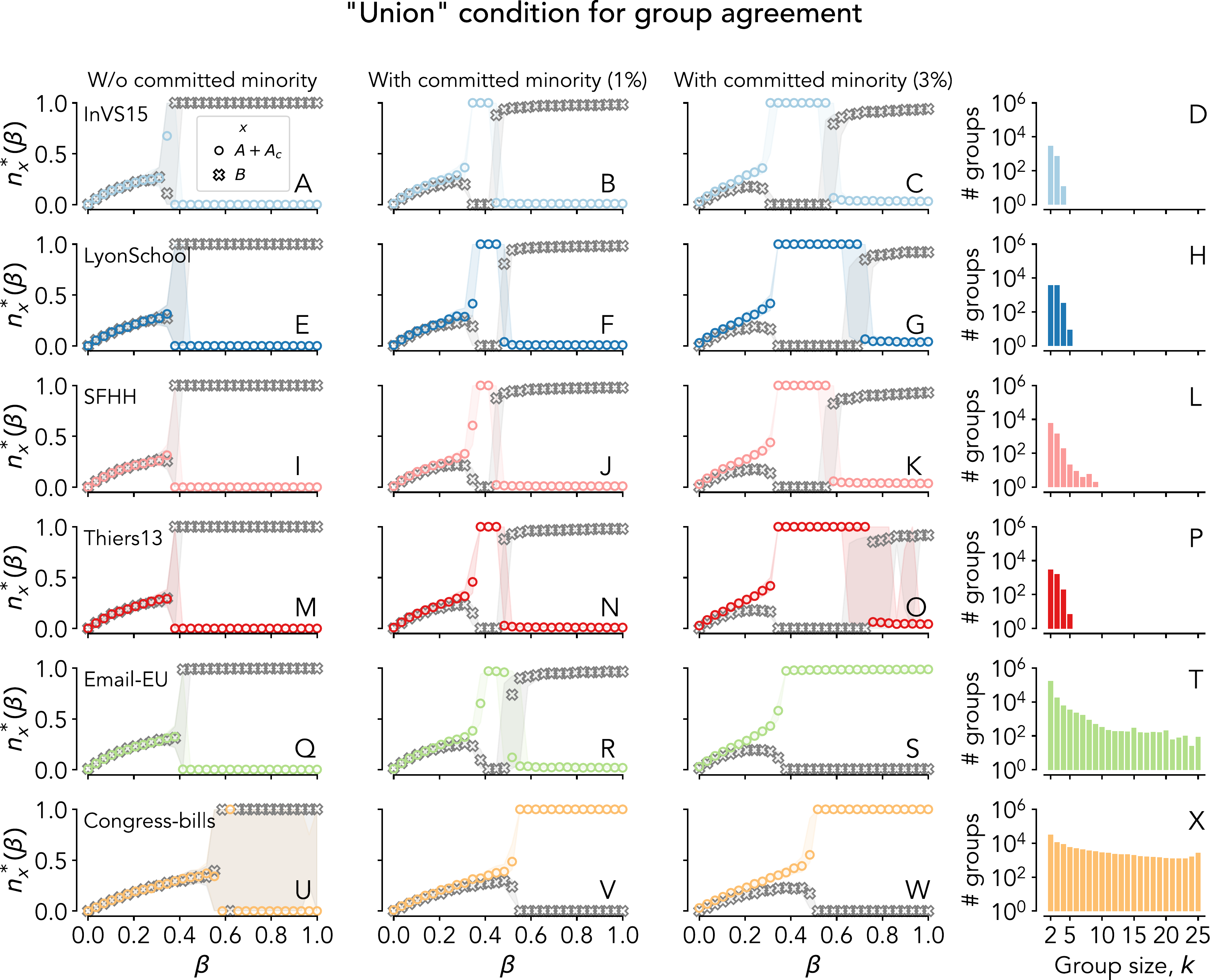}
	\caption{Simulations of the stationary state of the dynamics on empirical higher-order structures for a NG with the ``union'' condition for group agreement and associated group size frequencies.
		The considered social structures are the same of Fig.~2 of the main text and correspond to empirical data sets collected in six different context: workplace ({InVS15}), a primary school ({LyonSchool}), a conference ({SFHH}), a high school ({Thiers13}), email communications ({Email-EU}) and a political congress ({Congress-bills}). The phase diagrams in panels ({\it A,E,I,M,Q,U}), panels ({\it B,F,J,N,R,V}) and panels ({\it C, G, K, O, S, W}) correspond to simulation with a different fraction of committed minority, respectively set to $p=0$, $p=0.01$ and $p=0.03$. 
		The fraction of nodes supporting name $x$ in the stationary state $n^*_x(\beta)$, obtained by means of numerical simulations on each data set (row), is plotted as a function of the communication efficiency $\beta$.
		The results averaged over different runs of stochastic simulations are reported as points (circles and crosses, respectively associated to name $A$ and $B$) and shaded areas, representing median values and values contained within the 25th and 75th percentiles.
		Panels ({\it D,H,L,P,T,X}) show the group size frequencies associated to each data set, where a group of size $k$ represents a higher-order interaction between the $k$ nodes involved.
	}\label{fig:SI:sim_empdata_union}
\end{figure*}


\begin{figure}[]
	\centering
	\includegraphics[width=0.8\textwidth]{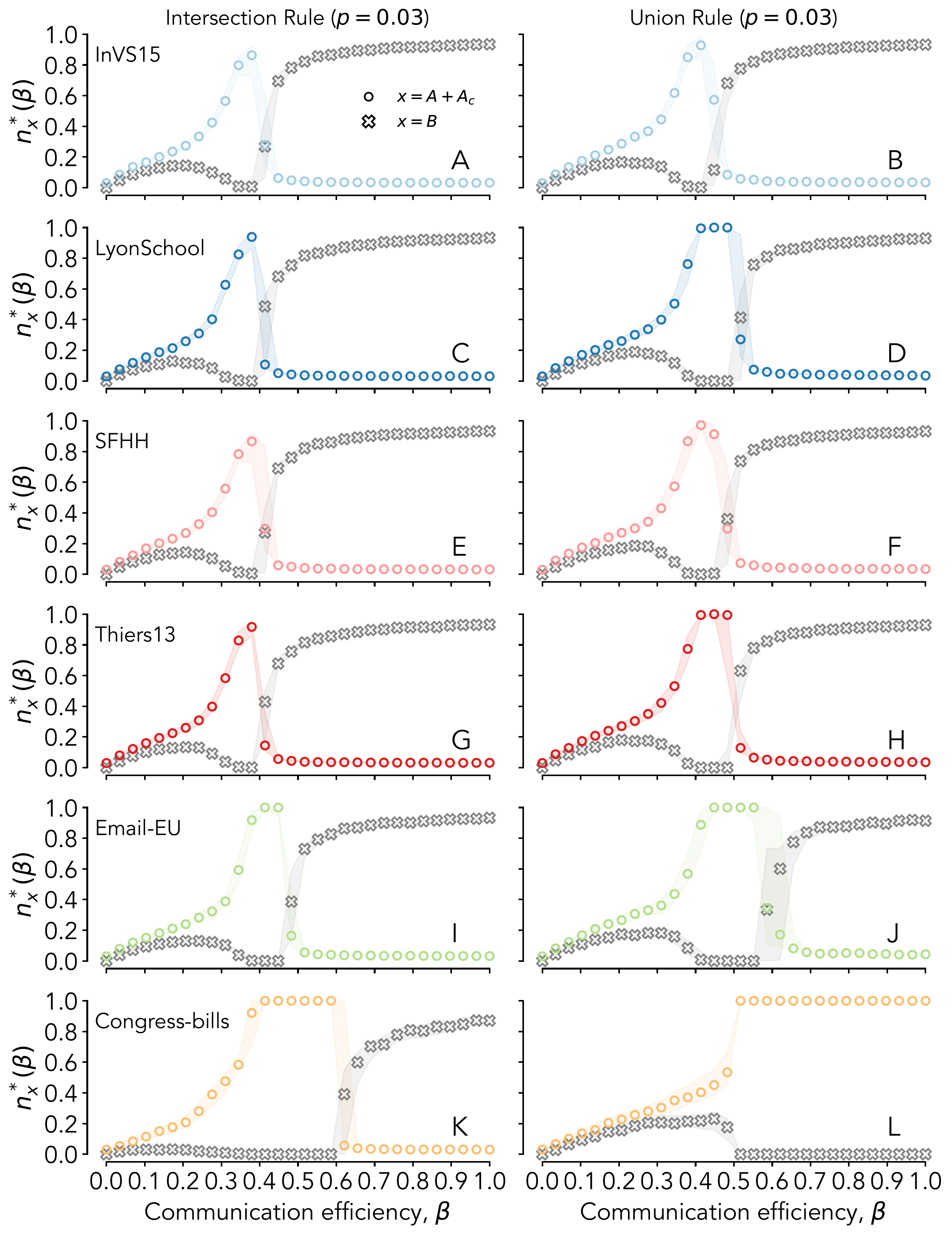}
	\caption{Group-size-based mean-field simulations of a NG on empirical higher-order structures constructed from data sets collected in six different context: workplace ({InVS15}), a primary school ({LyonSchool}), a conference ({SFHH}), a high school ({Thiers13}), email communications ({Email-EU}) and a political congress ({Congress-bills}). In all scenarios, simulations are initiated with a random selection of 3\% of committed minorities among the population.
		Left ({\it A,C,E,G,I,K}) and right ({\it B,D,F,H,J,L}) panels correspond to simulations respectively performed with the unanimity (intersection) and the union condition rule for group agreement. 
		Simulations are performed in the group-size-based mean-field approximation, that means that at each time step a size $k$ is sampled from the group size distribution of the correspondent data set, and then $k$ nodes are selected at random among the entire population.
		The fraction of nodes supporting name $x$ in the stationary state $n^*_x(\beta)$, obtained by means of 50 numerical simulations on each data set (row), is plotted as a function of the communication efficiency $\beta$. Points (circles and crosses) refer to the median values (respectively associated to name $A+A_c$ and $B$), with shaded areas representing the values contained within the 25th and 75th percentiles. As these simulations discard the correlations present in the original data, the results are only influenced by the group size distribution. In particular, the broad range on which the minority wins for the Congress-bills data set with the union rule can be attributed to the much broader distribution of group sizes with respect to the other data sets.
	}
	\label{fig:SI:sim_p_HomMix_empdata}
\end{figure}


\begin{figure}
	\centering
	\includegraphics[width=\textwidth]{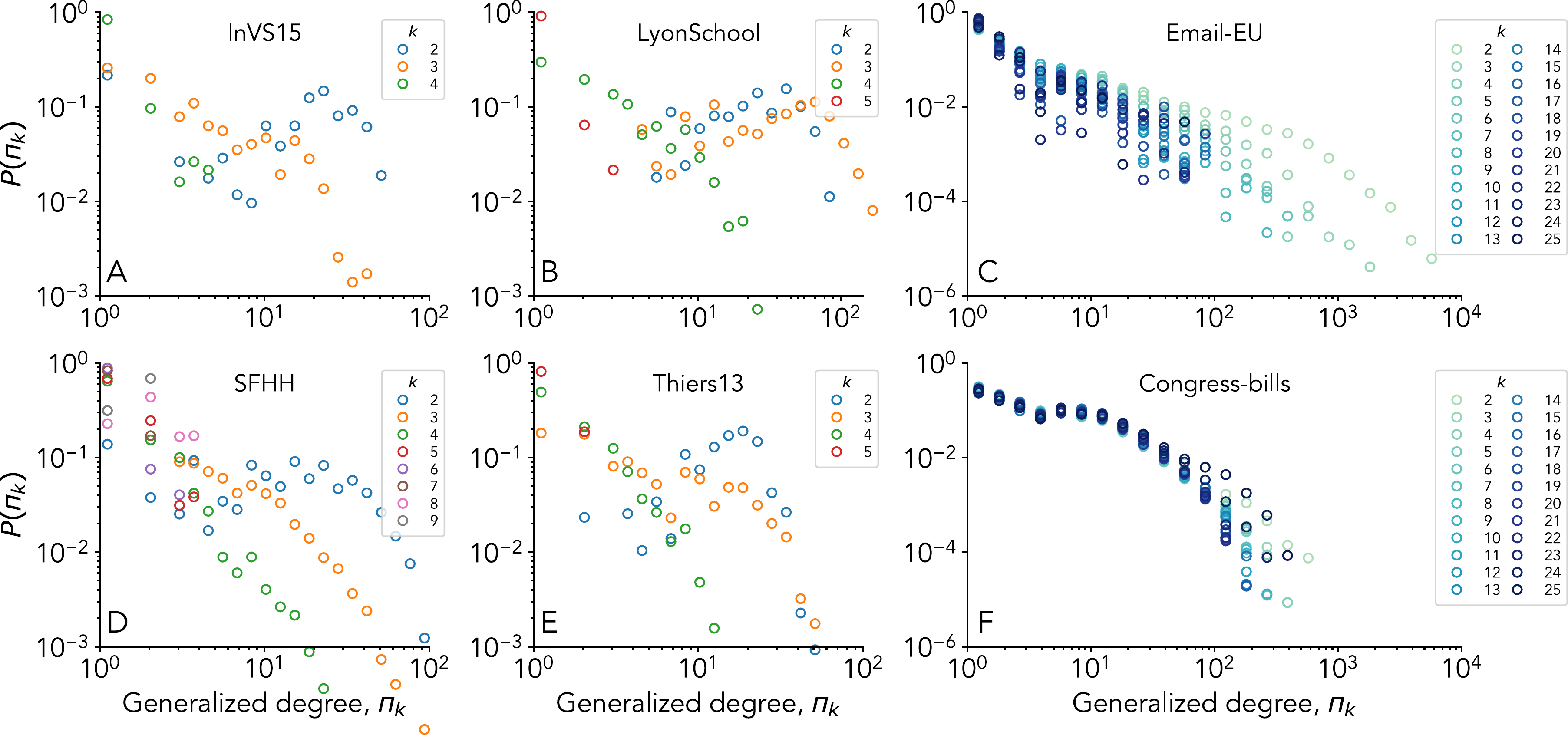}
	\caption{Generalised degree distributions of the considered empirical hypergraphs, each constructed from data sets collected in six different context: workplace ({InVS15}), a primary school ({LyonSchool}), a conference ({SFHH}), a high school ({Thiers13}), email communications ({Email-EU}) and a political congress ({Congress-bills}). The generalised degree $\pi_k$ of a node denotes the number of groups of size $k$ it is part of.}
	\label{fig:SI:gen_deg_dist}
\end{figure}


\begin{figure}
	\centering
	\includegraphics[width=\textwidth]{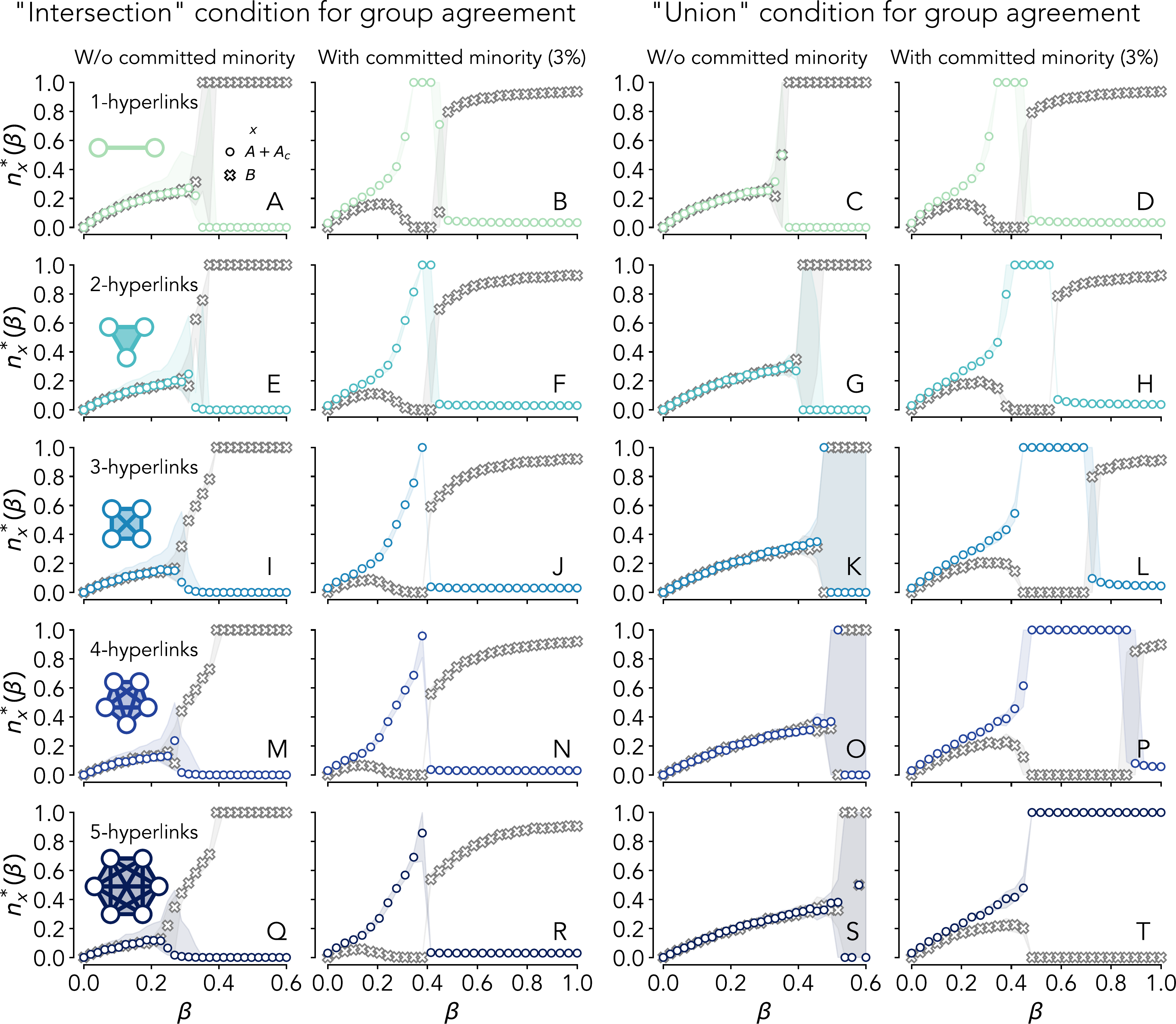}
	\caption{
		Higher-order (group) effects in NG with variable $\beta$.
		The fraction of nodes supporting name $x$ in the stationary state, $n^*_x(\beta)$, obtained by means of numerical simulations, is shown as a function of $\beta$. Lines (continuous and dashed, respectively associated to norms $A+A_c$ and $B$) and shaded areas refer to the median values and values contained within the 25th and 75th percentiles of the 50 numerical simulations. Panels in the first and third column correspond to simulation without committed minorities ($p=0$), while panels in the second and fourth column correspond to simulation with a non-zero fraction of committed minorities supporting name $A$ ($p=0.03$). Note that the scales on the x-axes vary for different values of $p$.
		Two different conditions for group agreement are considered: unanimity (``intersection'') rule (first two columns) and ``union'' rule (last two columns).
	}\label{fig:SI:sim_groupsize}
\end{figure}


\begin{figure}
	\centering
	\includegraphics[width=0.8\textwidth]{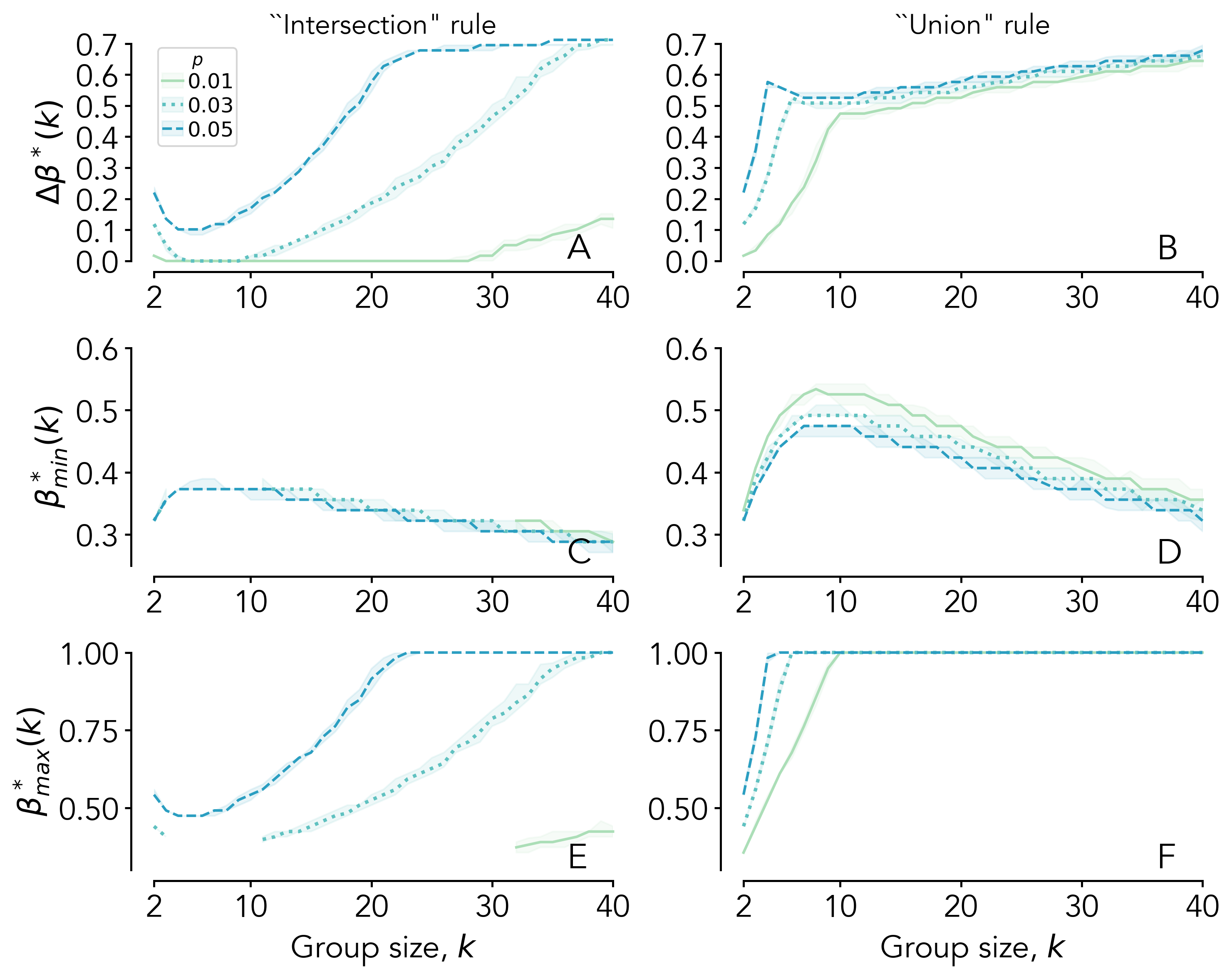}
	\caption{Higher-order (group) effects in NG for different values of group size $k$ and group agreement rule.
			Left ({\it A, C, E}) and right ({\it B, D, F}) panels correspond to simulations respectively performed with the unanimity (intersection) and the union condition for group agreement.
			We consider $(k-1)-$uniform hypergraphs, i.e. regular structures in which each interaction involves exactly $k$ agents. 
			({\it A, B}) The range $\Delta\beta^*$ of $\beta$ values for which $n^*_A=1$ (i.e., the committed minority manages to convert the whole population), is plotted as a function of the group size $k$ and for different values of $p$ (see legend). The associated minimal value $\beta^*_{min}$ and maximal value $\beta^*_{max}$ are respectively shown in panels ({\it C, D}) and ({\it E, F}).
	}
\end{figure}


\begin{figure}
	\centering
	\includegraphics[width=0.5\textwidth]{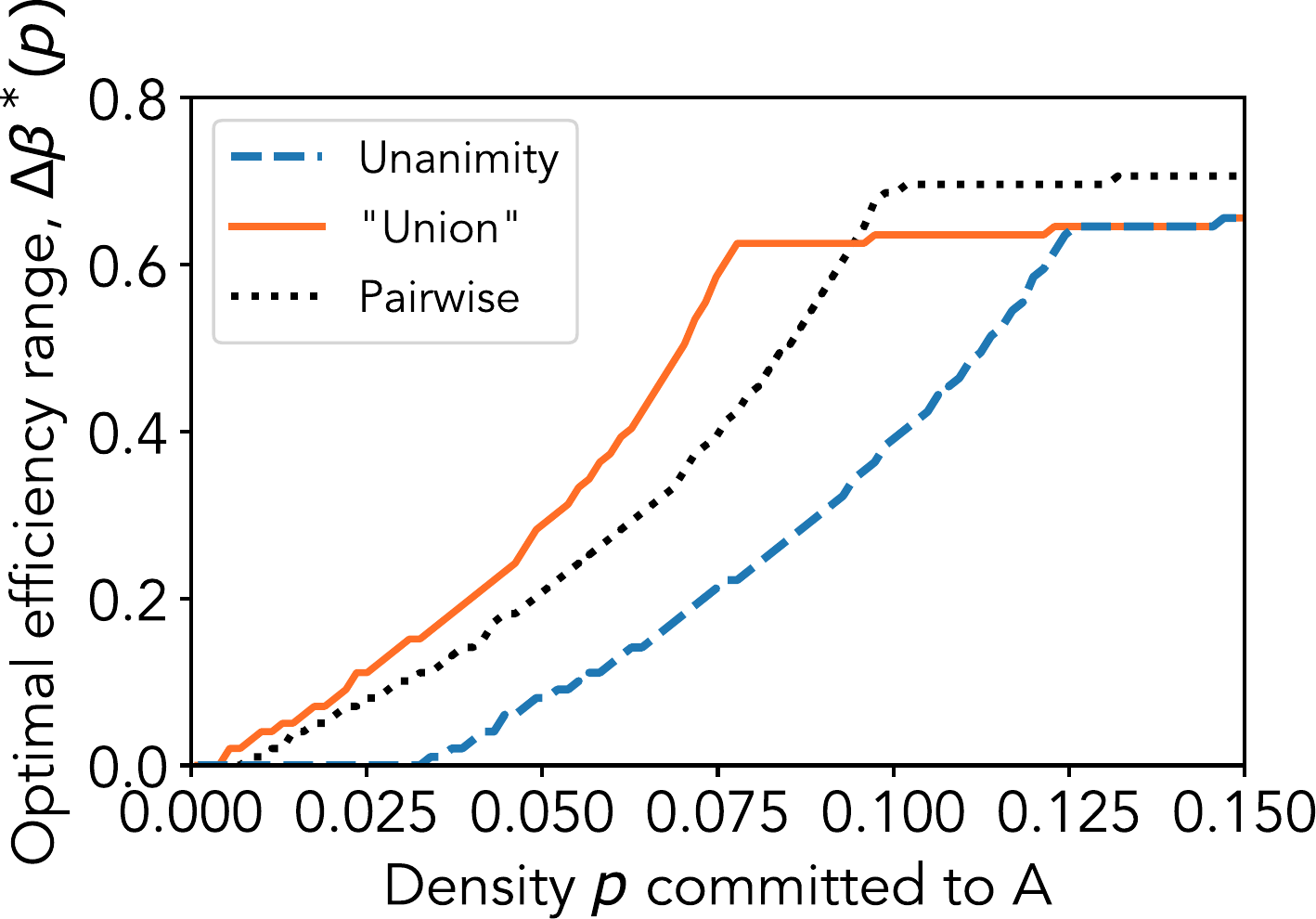}
	\caption{Mean field approximation on 2-uniform hypergraphs. The range $\Delta\beta^*$ of $\beta$ values for which $n^*_A=1$ (i.e., the committed minority manages to convert the whole population), is plotted as a function of the the fraction $p$ of agents committed to $A$ for different conditions for group agreement, namely unanimity ({\it A}) and ``union'' ({\it B}). For comparison, ({\it C}) corresponds to the pairwise version of the NG model where no group agreement condition can be defined.
		Values for are obtained through numerical integration of the correspondent system of equations, Eq.~\eqref{eq:SI:MF_int}, Eq.~\eqref{eq:SI:MF_union} and Eq.~\eqref{eq:SI:MF_pairwise}, after setting $p$, $n_A(0)=0$, $n_B(0)=1-p$, and $n_{AB}(0)=0$. 
	}
\end{figure}


\begin{figure}
	\centering
	\includegraphics[width=\textwidth]{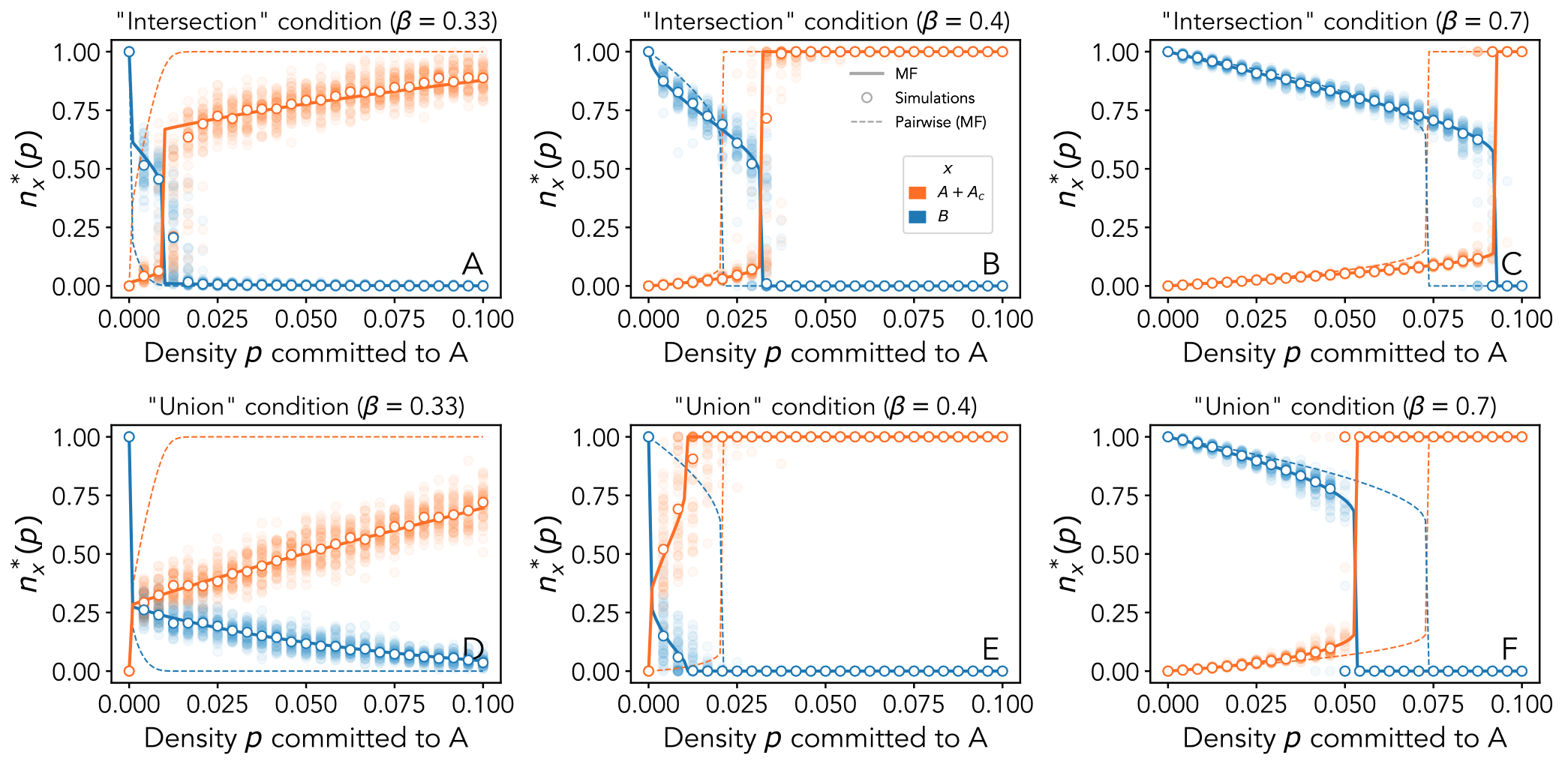}
	\caption{Testing the mean field approach against simulations.
		Stationary fraction $n^{*}_x(\beta)$ of agents supporting name $x$ as a function of the fraction $p>0$ of agents committed to $A$ for a NG with ``intersection'' ({\it A-C}) and ``union'' ({\it D-F}) conditions for group agreement.
		Different columns correspond to different values of $\beta$. Continuous lines are obtained through numerical integration of the mean field equations Eq.~\eqref{eq:SI:MF_int}, Eq.~\eqref{eq:SI:MF_union} after setting $p>0$, $n_A(0)=0$, $n_B(0)=1-p$, and $n_{AB}(0)=0$.
		Points are the results of Monte Carlo simulations. The pairwise case is plotted with dashed lines for comparison.
	}
	\label{fig:SI:MF_CM_n_vs_p}
\end{figure}

\begin{figure}
	\centering
	\includegraphics[width=\textwidth]{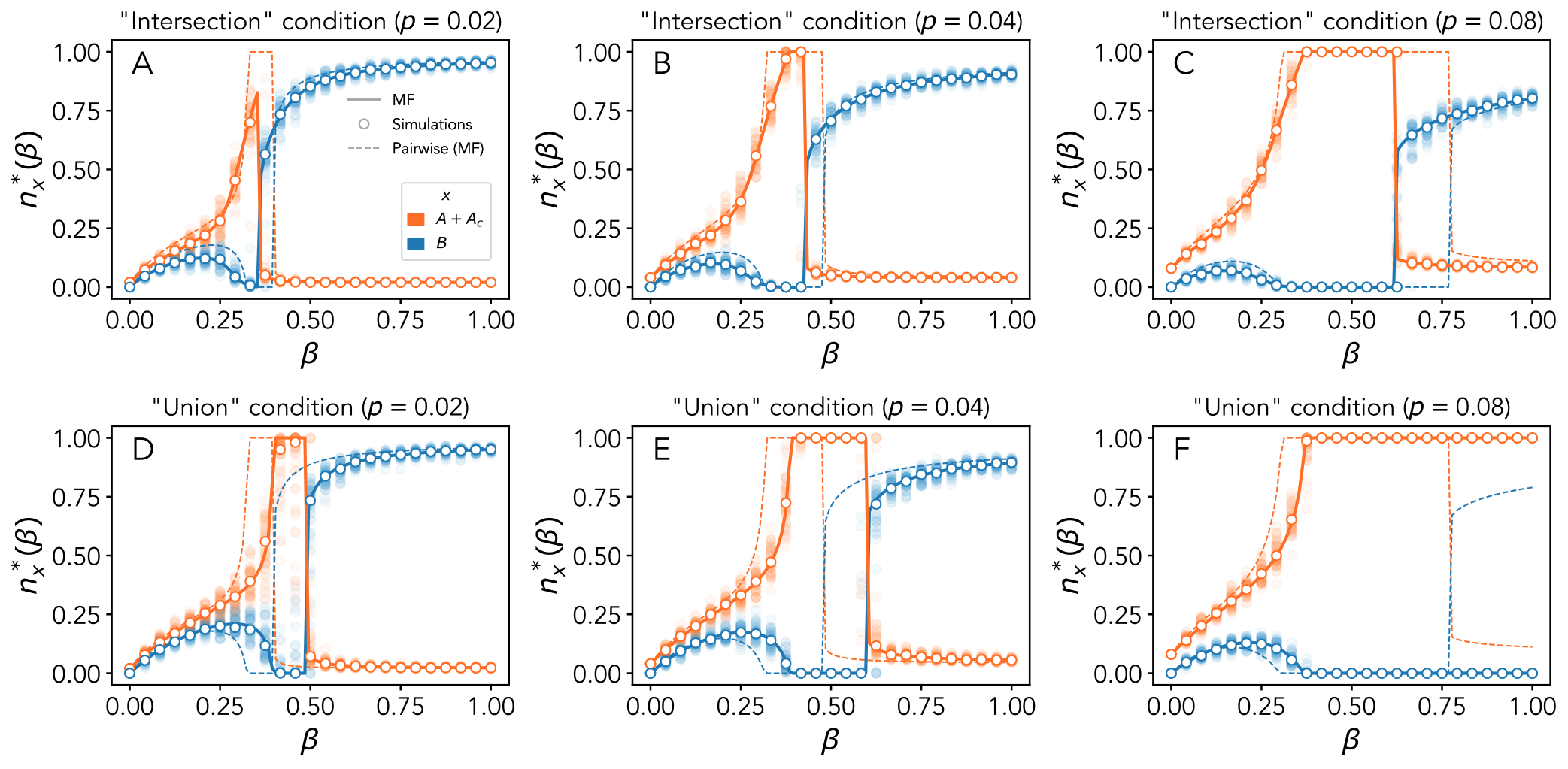}
	\caption{Testing the mean field approach against simulations.
		Stationary fraction $n^{*}_x(\beta)$ of agents supporting name $x$ as a function of the efficiency of social influence $\beta$ for a NG with ``intersection'' ({\it A-C}) and ``union'' ({\it D-F}) conditions for group agreement.
		Different columns correspond to different values of the fraction $p>0$ of agents committed to $A$ . Continuous lines are obtained through numerical integration of the mean field equations Eq.~\eqref{eq:SI:MF_int}, Eq.~\eqref{eq:SI:MF_union} after setting $p>0$, $n_A(0)=0$, $n_B(0)=1-p$, and $n_{AB}(0)=0$.
		Points are the results of Monte Carlo simulations. The pairwise case is plotted with dashed lines for comparison.}
	\label{fig:SI:MF_CM_n_vs_beta}
\end{figure}


\begin{figure}
	\centering
	\includegraphics[width=\textwidth]{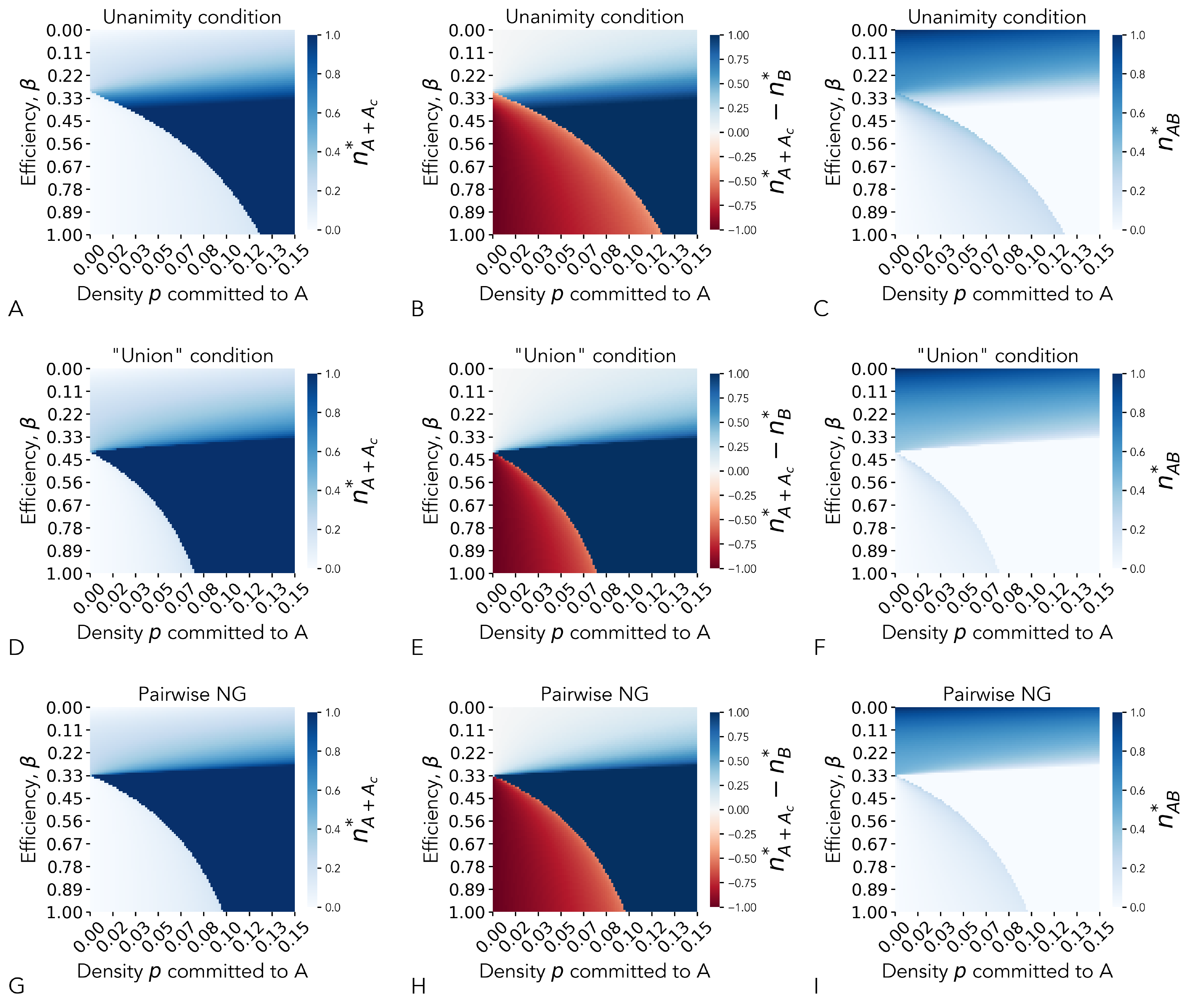}
	\caption{Two-dimensional phase diagrams of the NG in the mean field approximation.
		Heatmaps of the stationary fraction of agents supporting a certain name as a function of the efficiency of social infulence $\beta$ and the fraction $p$ of agents committed to name $A$ for unanimity ({\it A-C}) and ``union'' ({\it D-F}) condition for group agreement on 2-uniform hypergraphs. For comparison, the pairwise version of the NG model is reported in ({\it G-I}), where no group agreement condition can be defined.
		Values for each rule are obtained through numerical integration of the correspondent system of equations, Eq.~\eqref{eq:SI:MF_int}, Eq.~\eqref{eq:SI:MF_union} and Eq.~\eqref{eq:SI:MF_pairwise}, after setting $p$, $n_A(0)=0$, $n_B(0)=1-p$, and $n_{AB}(0)=0$. 
		Panels ({\it A, D, G}) show the stationary fraction of agents supporting name $A$, while its difference with respect to the ones supporting $B$ is displayed in ({\it B,E,H}). Finally, panels ({\it C,F,I}) show the stationary fraction of agents holding both names $A,B$.
	}\label{fig:SI:MF_heatmaps}
\end{figure}


\begin{figure}
	\centering
	\includegraphics[width=\textwidth]{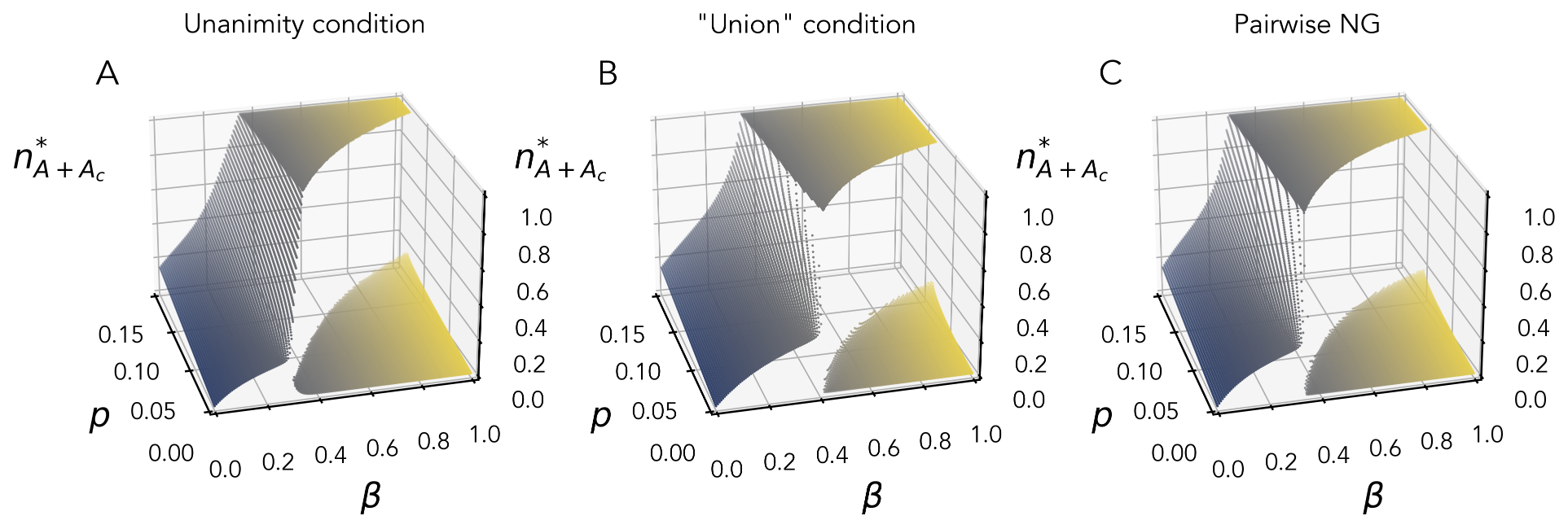}
	\caption{Three-dimensional phase diagram of the NG in the mean-field approximation.
		The stationary fraction of agents supporting name $A$, which includes committed agents, is plotted (z-axis) as a function of the efficiency of social influence $\beta$ and the fraction $p$ of agents committed to $A$. Different panels correspond to different conditions for group agreement, namely unanimity ({\it A}) and ``union'' ({\it B}), on 2-uniform hypergraphs. The pairwise version of the NG model is reported in ({\it C}) for comparison, where no group agreement condition can be defined.
		Values for each rule are obtained through numerical integration of the correspondent system of equations, namely Eq.~\eqref{eq:SI:MF_int}, Eq.~\eqref{eq:SI:MF_union} and Eq.~\eqref{eq:SI:MF_pairwise}, after setting $p$, $n_A(0)=0$, $n_B(0)=1-p$, and $n_{AB}(0)=0$. 
	}
	\label{fig:SI:MF_CM_3D}
\end{figure}


\end{document}